\documentclass[lettersize,journal]{IEEEtran}
\usepackage{amsmath,amsfonts,amsthm,amssymb}
\usepackage{amsmath,bm}
\usepackage{algorithmic}
\usepackage{algorithm}
\usepackage{array}
\newtheorem{lemma}{Lemma}
\newtheorem{theorem}{Theorem}
\newtheorem{remark}{Remark}
\usepackage[caption=false,font=normalsize,labelfont=scriptsize,textfont=sf]{subfig}
\usepackage{textcomp}
\usepackage{stfloats}
\usepackage{url}
\usepackage{verbatim}
\usepackage{graphicx}
\usepackage{cite}
\usepackage{tabularx}
\usepackage{capt-of}  
\begin{document}

\title{Latency-Distortion Tradeoffs in Communicating Classification Results over Noisy Channels}

\author{\IEEEauthorblockN{Noel Teku \quad Sudarshan Adiga \quad Ravi Tandon}\\
\IEEEauthorblockA{{Department of Electrical and Computer Engineering } \\
{University of Arizona, Tucson, AZ, USA} \\
Email: \{\emph{nteku1}, \emph{adiga}, \emph{tandonr}\}@arizona.edu}

\thanks{This work was supported by NSF grants CAREER 1651492, CCF-2100013,
CNS-2209951, CNS-1822071, CNS-2317192. Part of this paper will be presented at the 2024 IEEE International Conference on Communications \cite{b1}.

The authors are with the Department of Electrical and Computer Engineering, The University of Arizona, Tucson, AZ 85721 USA (e- mail: nteku1@arizona.edu; adiga@arizona.edu; tandonr@arizona.edu).}
}


\IEEEpubid{0000--0000/00\$00.00~\copyright~2024 IEEE}

\maketitle

\begin{abstract}
In this work, the problem of communicating decisions of a classifier over a noisy channel is considered. With machine learning based models being used in variety of time-sensitive applications, transmission of these decisions in a reliable and timely manner is of significant importance. To this end, we study the scenario where a probability vector (representing the  decisions of a classifier) at the transmitter, needs to be transmitted over a noisy channel. Assuming that the distortion between the original probability vector and the reconstructed one at the receiver is measured via f-divergence, we study the trade-off between transmission latency and the distortion. 
We completely analyze this trade-off using uniform, lattice, and sparse lattice-based quantization techniques to encode the probability vector by first characterizing bit budgets for each technique given a requirement on the allowed source distortion. These bounds are then combined with results from finite-blocklength literature to provide a framework for analyzing the effects of both quantization distortion and distortion due to decoding error probability (i.e., channel effects) on the incurred transmission latency. 
Our results show that there is an interesting interplay between source distortion (i.e., distortion for the probability vector measured via f-divergence) and the subsequent channel encoding/decoding parameters; and indicate that a \textit{joint} design of these parameters is crucial to navigate the latency-distortion tradeoff. We study the impact of changing different parameters (e.g. number of classes, SNR, source distortion) on the latency-distortion tradeoff and perform experiments on AWGN and fading channels. Our results indicate that sparse lattice-based quantization is the most effective at minimizing latency for low end-to-end distortion requirements across different parameters and works best for sparse, high-dimensional probability vectors (i.e., high number of classes).
\end{abstract}


\section{Introduction}
\IEEEPARstart{I}{n} recent years, machine learning (ML) has been increasingly applied to time-sensitive applications, including Vehicle to Vehicle (V2V) and Vehicle to Infrastructure (V2I) communications. These applications require reliable and rapid data transmission for tasks such as trajectory prediction \cite{b2} and lane change detection \cite{b3}. Similarly, this need for reliable, fast communication extends to other domains like internet of things (IoT) and edge computing. Coinciding with the increasing use of ML in low-latency applications, there has also been a growing body of work on context-dependent low-latency communications; which includes semantic communications  \cite{b4, b5, b6, b7},
 ultra-reliable low latency communications (URLLC) \cite{b8, b9}, and joint source channel coding \cite{b10,b11, b12, b13}. 

 Semantic communication  generally focuses on sending \textit{context dependent} features/decisions dependent on the data to the receiver (rather than the entire raw message) \cite{b7}. In doing so, the amount of bits required for transmission is often reduced \cite{b6}. For example, in \cite{b14}, a transformer-based network
was used to learn/transmit semantic features of sentences
and decode the received features to ensure that the original
meaning of the sentences were preserved. In \cite{b15}, an approach to modeling the length of a semantic message and its distortion based on noise due to the model and the channel is presented along with masking strategies that can be applied before transmission. \cite{b16, b17} present rate-distortion approaches for semantic communications for general block-wise distortion functions. The focus of URLLC is to design protocols in order to transmit low-data rate (short packets) with high reliability (low probability of error) within a small latency \cite{b9}. A rate-distortion analysis is also performed in \cite{b18} for short control packets, assuming transmissions are being made to a remote agent, where the distortion measures considered are quantization error and the freshness of the data (age of information); however, this analysis is done under the assumption of noiseless channels. 
   \begin{figure*}[!t]
	\begin{center}
		\includegraphics[scale = 0.28]{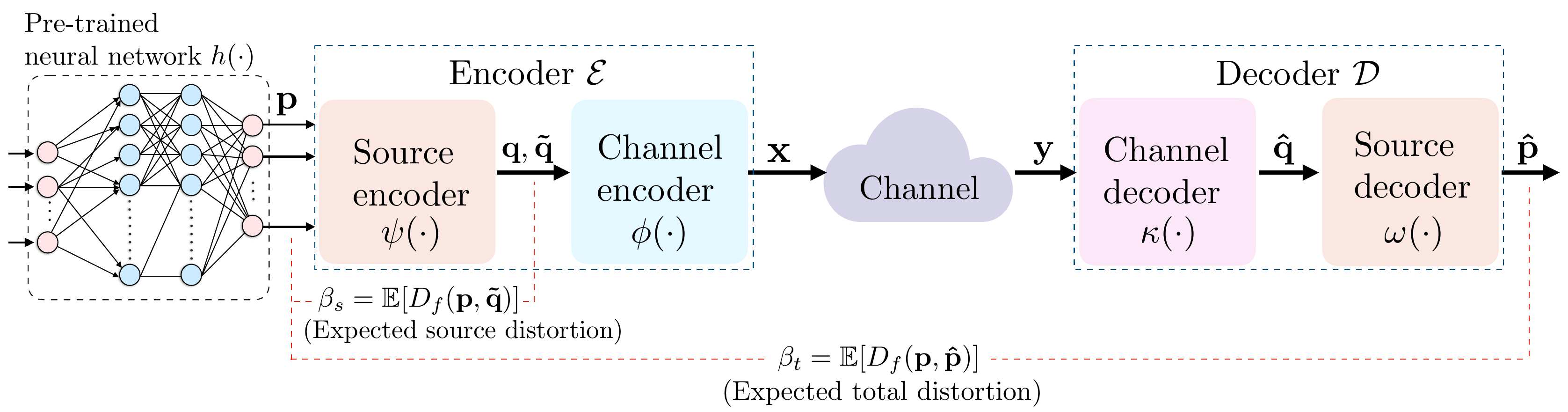}
	\caption{End-to-End block diagram for communicating classifier decisions (probability vector) over a noisy channel.}
 \label{fig1}
 \vspace{-0.55cm}
		\end{center}
\end{figure*}
\IEEEpubidadjcol

\underline{\textit{Overview and Main Contributions:}}  In this paper, we focus on the following problem: a transmitter wishes to send a probability vector (e.g., representing the decisions of a ML based classifier) to a receiver over a noisy channel. Our objective in focusing on transmitting probability vectors is to observe if there are any properties of such a vector that can be exploited for obtaining reductions in latency when transmitting while still preserving the decision of the classifier. When using datasets with a high number of classes (e.g. CIFAR-100, Imagenet-1k), for example, there will be many entries in the vector that do not add any substantial information regarding the classifier's decision. In such a scenario, a vector containing a small number of a classifier's highest predictions may be sufficient for identifying a classifier's decision compared to sending the entire vector. Results obtained from classifiers operating on real datasets are provided later in the paper that further support this idea. Additionally, transmitting probability vectors enables the use of specific quantization techniques that can be efficient. One such technique, which is considered in this work, is a lattice based quantization proposed in \cite{b19}, where a given probability distribution is fitted to its nearest match on a certain finite-dimensional lattice. This problem can also be viewed within the umbrella of semantic communication and joint source channel coding. Transmitting the results of a classification task incurs lower latency/overhead compared to sending a compressed form of the
data required for classification at the receiver. It also enables
the receiver to quickly execute tasks that depend on knowing
the classification results, which is essential to conducting goal-oriented communications \cite{b5}. Additionally, this problem falls under the umbrella of JSCC as its objective is to attain low-latency transmissions by operating in the finite blocklength regime \cite{b11}. 
\IEEEpubidadjcol

The main new elements herein are two fold: we measure utility of the reconstruction of the probability vector in terms of statistical divergence measures; and secondly, we simultaneously want to minimize the transmission latency over the noisy channel. We note that there has been prior work on quantizing probability distributions, including \cite{b19, b20, b21, b22,b23}.
 In particular, \cite{b22, b23} 
 investigated quantizing probability distributions in order to minimize  Kullback-Leibler (KL)-divergence by performing a non-linear operation and then using uniform quantization. However, the existing works did not study the scenario when a probability vector has to be transmitted through a noisy channel, and what would be the right quantization strategy/parameters if the goal is to minimize latency. By considering the distortion introduced by the channel and quantization, we aim to analyze the trade off between the end-to-end distortion of the system and the incurred transmission latency. Additionally, a similar framework considering quantization noise was introduced in \cite{b24}, but focused on transmitting control signals and ensuring the stability of the assumed control system rather than quantizing/transmitting probability vectors. There have also been works such as \cite{b25,b26,b27} that look at relating finite blocklength analysis with latencies but do not consider quantization noise. 

Our main contributions are as follows: 
\begin{itemize}
     \item \textit{In-depth investigation of quantization techniques for classification results:} The performance of uniform, lattice, and sparse lattice-based quantization techniques are investigated with respect to balancing the trade-off between latency and end-to-end distortion. We show that the lattice-based methods are more efficient than the baseline uniform quantization as they require less complexity and make use of the properties of the probability vector to require fewer bits. The sparse-lattice based technique is proposed to employ the assumed lattice-based quantization technique on only a few of the highest probabilities of the vector. This amount should be determined such that a large portion of the mass of the vector is represented, which is investigated on predictions from classifiers on real datasets; specifically, CIFAR-100 \& Imagenet-1K. 
     We provide results bounding the necessary bit budgets under each technique to satisfy a requirement on the allowable source distortion (Lemmas \ref{lma2}-\ref{lma4}). Our results show the expected trend that to ensure a lower source distortion when quantizing the probability vector, a higher bit budget is needed for each of the assumed techniques. For a probability vector of length $50$ classes and the same source distortion, for example, our results show that sparse-lattice based quantization incurs a reduction in bit requirement of approximately $96\%$ and $80\%$ with respect to uniform and lattice-based quantization.  
     \item \textit{Latency-Distortion trade-off analysis:}  
     We derive a relationship (Lemma \ref{lma5}) between the source distortion incurred for each of the afroementioned quantization techniques and the decoding error probability (i.e. accounting for distortion caused by noisy channel effects) to obtain a bound on the end-to-end distortion between the received and transmitted vectors. By incorporating these two sources of distortion, the subsequent blocklength under these parameters can be obtained and used to calculate the transmission latency (Theorem \ref{thm1}). Our results show that by using the proposed framework, an optimized source distortion can be found that achieves a minimal latency for different levels of end-to-end distortion. In doing so, this also enables us to extend our framework to fading channels (Theorems \ref{thm2} \& \ref{thm3}).
    \item \textit{Application to noisy channels:} We provide a comprehensive set of simulation results to validate the proposed framework. Specifically, we study the trade-off between latency and distortion while varying parameters of the framework; such as, channel conditions (i.e. SNR), source distortion, and the length of the probability vector (i.e. number of classes). We report results assuming additive white gaussian noise (AWGN) and fading channels using results from the literature on finite blocklength. For a probability vector of length $100$ classes and the same end-to-end distortion, for example, our results show that the sparse-lattice based quantization can incur a latency reduction of approximately $97\%$ and $85\%$ with respect to uniform and standard lattice-based quantization for the AWGN channel. Our results indicate that sparse lattice-based quantization is the most effective at minimizing latency for low end-to-end distortion requirements across different parameters. Specifically, the results indicate that sparse lattice-based quantization works best for sparse, high-dimensional probability vectors (i.e. high number of classes). 
\end{itemize}


The paper is structured as follows: Section \ref{sec:System_Model} presents the system model studied in this work; Section \ref{sec:Main} presents results analyzing the distortion incurred with each of the quantization techniques and details our framework for analyzing the latency-distortion tradeoff using these techniques for AWGN and fading channels; Section \ref{sec:AA} presents results from simulations; Section \ref{sec:conc} concludes the paper and proposes future work. The proofs for the technical results are presented in the Appendix.

\section{System Model}
\label{sec:System_Model}
We consider the scenario illustrated in Fig. \ref{fig1}: a pre-trained classifier (e.g., a neural network), denoted as $h(\cdot)$, is used for a $k$-class classification problem and is situated at a transmitter. The output classification probabilities are represented as $\mathbf{p} = [\mathbf{p}[1], \mathbf{p}[2], \cdots, \mathbf{p}[k]]^{\top}$, where $\mathbf{p} \in \mathbb{R}^{k \times 1}$.
Let $\mathbf{\hat{p}} = [\mathbf{\hat{p}}[1], \mathbf{\hat{p}}[2], \cdots, \mathbf{\hat{p}}[k]]^{\top}$ denote the estimated classifier output at the receiver. 
In this paper, we measure the distortion between $\mathbf{p}$ and $\mathbf{\hat{p}}$ via $f$-divergence, defined as 
\begin{align}
D_\text{f}(\mathbf{p},\mathbf{\hat{p}}) = \sum_{i=1}^{k}f\left(\frac{\mathbf{p}[i]}{\mathbf{\hat{p}}[i]}\right)\mathbf{\hat{p}}[i].
\end{align}
The transmitter's goal is to communicate the probability vector $\mathbf{p}$ within a latency budget of $T_{\text{max}}$ with minimum total expected distortion $\beta_t$, i.e., $\mathbb{E}(D_\text{f}(\mathbf{p},\mathbf{\hat{p}}))\leq \beta_t$, where the expectation is over the noisy channel realizations. 
We next describe the main components (source/channel encoder/decoder(s)): 
a source encoder $\psi(\cdot)$  quantizes the probability vector $\mathbf{p}$, such that $\mathbf{q} = \psi(\mathbf{p})$. The lossy compression caused by quantization results in source distortion, denoted by $\beta_s$. The total number of bits required by $\mathbf{q}$, given the source distortion, is represented as  $J(\beta_s)$, where $J(\cdot)$ is a function of $\beta_s$.  We note that based on the quantization technique, $\mathbf{q}$ may not necessarily be a probability vector. In this scenario, we normalize the values in $\mathbf{q}$ to obtain the corresponding probability vector $\mathbf{\tilde{q}}$ after source encoding, where $\mathbf{\tilde{q}}[i] = \mathbf{{q}}[i]/ \sum_{i=1}^{k}\mathbf{{q}}[i]$ and $\mathbf{\tilde{q}} \in \mathbb{R}^{k \times 1}$. The source distortion $\beta_s$ is quantified as $\beta_s = D_\text{f}(\mathbf{p},\mathbf{\tilde{q}})$. We use the channel encoder $\phi(\cdot)$ to generate the $n$-length channel input $\mathbf{x} = \phi(\mathbf{q})$, where $\mathbf{x} = [\mathbf{x}[1], \mathbf{x}[2], \cdots, \mathbf{x}[n]]^{\top}$ and $\mathbf{x} \in \mathcal{X}^{n}$. Let $\mathcal{E}$ denote the source and channel encoder pair. We consider a bandwidth constrained fading channel, where the channel output is given by $\mathbf{y}[i] = \mathbf{h}[i]\mathbf{x}[i] + \mathbf{z}[i]$, for all $i \in [n]$; where $\mathbf{y} \in \mathbb{R}^{n \times 1}$, the channel fading gains are given by  $\mathbf{h} = [\mathbf{h}[1],\mathbf{h}[2], \cdots, \mathbf{h}[n]]^{\top}$ with $\mathbf{h} \in \mathbb{R}^{n \times 1}$, and the AWGN noise vector is given by $\mathbf{z} = [\mathbf{z}[1],\mathbf{z}[2], \cdots, \mathbf{z}[n]]^{\top}$ with $\mathbf{z} \in \mathbb{R}^{n \times 1}$. We also consider a bandwidth constrained AWGN channel which can be obtained from the above model by setting $\mathbf{h}[i] = 1 \ \forall \ i \in n$. The signal-to-noise ratio (SNR) of the channel for a bandwidth $B_0$ Hz, is defined as $\gamma_0 = \frac{P}{N_0}$, where $P$ denotes the signal power and $N_0$ denotes the noise power. To simulate using the same transmit powers at different bandwidths, as done in \cite{b8}, we define the operational SNR for a channel of bandwidth $B$ Hz as $\gamma = \frac{\gamma_0B_0}{B}$   where $\frac{B_0}{B}$ acts as a scaling factor for relating different channel conditions.

We denote the decoding error probability by $\epsilon^{*}(n)$, where $\epsilon^{*}(n) \in [0,1]$. At the receiver, we consider a channel decoder, denoted by $\kappa(\cdot)$, such that $\mathbf{\hat{q}} = \kappa(\mathbf{y})$. Subsequently, we consider the source decoder $\omega(\cdot)$ and a normalization operation to obtain an estimate of the classifier probabilities, given by $\mathbf{\hat{p}} = \omega(\mathbf{\hat{q}})$. Let $\mathcal{D}$ denote the source and channel decoder pair. 

The channel noise, in addition to the source distortion, contributes to the total end-to-end distortion. Given a specific SNR, it is possible to vary the source distortion $\beta_\text{s}$ to achieve a maximum total expected distortion of $\beta_\text{t}$. In other words, we have $\beta_s \in [0,\beta_\text{t}]$. This choice will also affect the incurred transmission latency; given a bandwidth of $B$ Hz, the time required to transmit an $n$-length vector $\mathbf{x}$ is calculated as:
\begin{align}
\label{eq:latency}
    T(\mathcal{E}, \mathcal{D}) = \frac{n}{2B}.
\end{align}

In this paper, we focus on understanding the tradeoff between latency and distortion for the task of communicating probability distributions. 
Specifically, given the channel statistics (e.g., bandwidth, SNR) and desired maximum latency $T_\text{max}$, the optimal distortion can be defined as follows:

\begin{align}
\label{eq:opti2}
D^{*}(T_{\text{max}})\triangleq \min_{(\mathcal{E},\mathcal{D})}\quad & \beta_t (\mathcal{E},\mathcal{D}), 
\textrm{~~s.t.~~}  T(\mathcal{E},\mathcal{D}) \leq T_{\text{max}}. 
\end{align}
Alternatively, we can fix the maximum permissible distortion $\beta_{\text{max}}$, and minimize the total latency $T$ over encoder-decoder pairs as
\begin{align}
T^{*}(\beta_{\text{max}}) \triangleq  \underset{(\mathcal{E},\mathcal{D})}{\min}\quad & T(\mathcal{E},\mathcal{D}), \textrm{~~s.t.~~}  \beta_t(\mathcal{E},\mathcal{D}) \leq \beta_{\text{max}}.  
\end{align}
In the lemma stated next (proof is presented in the Appendix), we show that the optimal latency $T^{*}(\beta_{\text{max}})$ is a convex non-increasing function of the total distortion $\beta_{\text{max}}$; and likewise, we show that the minimal distortion $D^{*}(T_{\text{max}})$ is a convex non-increasing function of $T_{\text{max}}$.

\begin{lemma}\label{lma1}
$T^{*}(\beta_{\text{max}})$ is convex non-increasing function of $\beta_{\text{max}}$.
$D^{*}(T_\text{max})$ is convex non-increasing function of $T_\text{max}$.
\end{lemma}

\section{Main Results \& Discussion}
\label{sec:Main}
In this section, we present the framework for analyzing the latency-distortion tradeoff. 
We begin by assuming a noiseless channel and uniform quantization as the source encoder (i.e., transforming $\mathbf{p}$ to $\mathbf{q}$) and analyze the corresponding source distortion (Lemma \ref{lma2}). We perform a similar analysis for lattice and sparse lattice-based quantization techniques to analyze the source distortion for a noiseless channel (Lemma \ref{lma3} \& Lemma \ref{lma4}). We then incorporate and analyze the impact of channel noise on the end-to-end distortion (Lemma \ref{lma5}). Subsequently, we use results on finite-blocklength capacity, which allow us to connect latency with the overall distortion. This, in turn, also leads to an explicit optimization (Theorem \ref{thm1}), which can be solved to  trade latency with distortion. We then extend this result to account for fading channels with and without CSI (Theorems \ref{thm2} \& \ref{thm3}). 
 \vspace{-4pt}
\subsection{Quantizing Classifier Probabilities}
\label{sec:quant}
\noindent\subsubsection{Uniform Quantization (UQ)}
Suppose we have a total budget of  $J$ bits to  quantize the $k$-dimensional probability vector $\mathbf{p}$. Under uniform quantization (UQ), we use $j= \lfloor{J/k}\rfloor$ bits to quantize each element $\mathbf{p}[i], i=1,2,\ldots,k$. We denote $\mathbf{q}[i]$ as the resulting quantized output. Note that $\mathbf{q}$ may not necessarily be a probability vector. We can however, normalize it as $\mathbf{\tilde{q}}[i]= \frac{\mathbf{q}[i]}{\sum_{\ell=1}^{k}\mathbf{q}[\ell]}$, for $i=1,\ldots, k$.   
Our objective is to minimize the $f$-divergence between $\mathbf{p}$ \& $\hat{\mathbf{p}}$; in the noiseless scenario, which would be equivalent to minimizing $D_\text{f}(\mathbf{p},{\mathbf{\tilde{q}}})$, as $\beta_s$ would be the only distortion present. When $f(x) = \frac{1}{2}|x-1|$, $f$-divergence results in total variation (TV): $D_\text{f}(\mathbf{p},{\mathbf{q}}) = D_\text{TV}(\mathbf{p},{\mathbf{q}}) = \frac{1}{2}\sum_{i=1}^{K}|\mathbf{p}[i]-{\mathbf{q}}[i]|$\cite{b28}.
The next lemma shows a sufficient condition on the quantization budget to achieve a source distortion of $\beta_s$.

 \begin{lemma}
 \noindent
 \label{lma2}
 For a $k$-class classification problem, if the total uniform quantization (UQ) budget satisfies 
 \vspace{-4pt}
 \begin{align}
 \label{eq:unifbits}
 J_{\text{UQ}}  \geq 2k \cdot \log_2 \left(\frac{k}{\beta_s}\right),
\end{align}
  \vspace{-4pt}
 then $ D_\text{TV}(\mathbf{p},\mathbf{\tilde{q}})\leq \beta_s$.  
\end{lemma}

\begin{remark} \label{rem1} (Impact of normalization)
The proof of Lemma \ref{lma2} is non-trivial due to the nature of the vectors involved. While $\mathbf{p}$ is a probability vector, the corresponding quantized $\mathbf{q}$ may not be a probability vector. To apply the statistical $f$-divergence measure, we normalize the entries of $\mathbf{q}$ by their sum, i.e., $S= \sum_{i}\mathbf{q}[i]$. However, this normalization operation makes the analysis of bounding the f-divergence challenging. The proof above overcomes this issue, by first assuming that the number of bits $j$ is of the form $j= \log(k/2\alpha)$, and then we are able to bound the sum $S$ as $S \in [1-\alpha, 1+\alpha]$. This allows us to determine the number of bits required to achieve a desired source distortion $\beta_s$. 
\end{remark}

\subsubsection{Lattice-based Quantization (LQ)} There are a few disadvantages when using UQ. First, UQ does not exploit the fact that the vector being compressed is a probability distribution. There are more efficient methods that can further reduce the number of bits required for quantization by exploiting this property.
Additionally, as the number of classes $k$ increases, the length of $\mathbf{p}$ will increase, leading to a significant increase in the number of bits required to satisfy the source distortion requirement as shown in Lemma $\ref{lma2}$. Also, as noted in Remark \ref{rem1}, UQ requires an additional normalization step which complicates deriving a bound on the source distortion between $\mathbf{p}$ and $\tilde{\mathbf{q}}$. Subsequently, we now consider the algorithm presented in \cite{b19}, which presents a lattice-based approach for quantizing probability distributions. The algorithm uses a lattice to represent a set of $k$-length probability distributions that is a subset of the $k$-dimensional probability simplex (i.e. the set of all possible $k$-length probability vectors $\mathcal{A}_k =\{[\mathbf{q}[1],...,\mathbf{q}[k]]\in \mathbb{Q}^{k} \mid \sum_i \mathbf{q}[i] = 1\}$). The probability vectors in the lattice are defined to have the property that each of their elements must have the same denominator $\ell$, which is a positive integer set by the user. Subsequently, each element in the probability vector must be of the form $\mathbf{q}[i] = \frac{\mathbf{b}[i]}{\ell}$, where $\mathbf{b}[i]$ are also positive integers. Because $\mathbf{q}$ must sum to 1, this implies that $\sum_i \mathbf{b}[i] = \ell$. Denoting the lattice as $Q_\ell$, the formal structure for the lattice is given as follows \cite{b19}:
\begin{equation}
\label{eq:lattice}
    Q_\ell = \{[\mathbf{q}[1],...,\mathbf{q}[k]]\in \mathbb{Q}^{k} \mid \mathbf{q}[i] = \frac{\mathbf{b}[i]}{\ell}, \sum_i \mathbf{b}[i] = \ell\}.
\end{equation}

\begin{algorithm}[t]
\caption{Lattice-based Quantization (LQ) \cite{b19}}\label{alg:alg1}
\begin{algorithmic}
\STATE \textbf{Inputs:} \textbf{p}, $Q_\ell$
\STATE Compute $\textbf{b}^{'}[i]= \lfloor \ell\textbf{p}[i] +\frac{1}{2} \rfloor, \ell' = \sum_i \textbf{b}^{'}[i]$
\IF{$\ell^{'}=\ell$}
\STATE Done
\ELSE
\STATE Calculate $\zeta[i] = \textbf{b}^{'}[i] - \ell\textbf{p}[i]$ and sort in increasing order.
\IF{$\ell^{'}-\ell >0$}
\STATE Decrease $|\ell'-\ell|$ values with largest $\zeta[i]$ in $\textbf{b}^{'}[i]$ by 1
\ELSIF{$\ell'-\ell<0$}
\STATE Increase $|\ell'-\ell|$ values with smallest $\zeta[i]$ in $\textbf{b}^{'}[i]$ by 1
\ENDIF
\ENDIF
\STATE Compute lexicographic index to represent $\textbf{b}[1],..., \textbf{b}[k]$
\end{algorithmic}
\label{alg1}
\end{algorithm}
\vspace{-5pt}

From \eqref{eq:lattice}, we can see that $Q_\ell \subseteq \mathcal{A}_k$ and if a probability distribution satisfies \eqref{eq:lattice}, it is a point on $Q_\ell$. The algorithm's objective is to find the point (i.e. probability distribution) on $Q_\ell$ closest, under an assumed distance metric, to a given probability distribution $\mathbf{p}$. We denote the resulting probability distribution chosen from $Q_\ell$ as $\mathbf{q}_\text{LQ}(\mathbf{p})$. The procedure for this method is summarized in Algorithm \ref{alg:alg1}. First, an initial guess of the nearest distribution based on $\mathbf{p}$ and $\ell$ is made using a simple mapping. If the mapping immediately results in a probability distribution on $Q_\ell$ the algorithm is complete; otherwise, updates are made to the guess based on the observed error to push it to the nearest point on $Q_\ell$. Thus, by mapping $\mathbf{p}$ to one of the available distributions on $Q_\ell$, lattice-based quantization (LQ) requires significantly less complexity compared to UQ. As an example, assume that we are given a probability vector $\mathbf{p} = [0.18, 0.52, 0.3]$ (meaning that $k=3$) and $\ell=5$. This means that $Q_5$ consists of all probability vectors of length $3$ whose entries have $5$ as a denominator. Examples of candidate probability vectors in $Q_5$ include [$\frac{1}{5}$,$\frac{2}{5}$,$\frac{2}{5}$],  [$\frac{1}{5}$,$\frac{1}{5}$,$\frac{2}{5}$], and [$\frac{3}{5}$,$\frac{2}{5}$,$0$]. Applying the initial mapping of $\mathbf{p}$ as shown in Algorithm \ref{alg:alg1}, results in an initial guess of $\mathbf{b}' = [1,3,2]$. However, this means that we have $\ell' = \sum_i \mathbf{b}'[i] = 6$. Because $\ell' \neq \ell$, we must perform updates to push this guess closer to $Q_\ell$. We first calculate how far away the guess is from an actual point on $Q_\ell$ by performing $\zeta[i] = \mathbf{b}[i] - \ell\mathbf{p}[i]$, which results in $\zeta = [0.1,0.4,0.5]$. Because $\ell'-\ell = 1$, we must decrement the element in $\mathbf{b}'$ with the largest $\zeta[i]$ by 1. From this example, we can see that the third element in $\mathbf{b}'$ has the largest error; decrementing it gives us $\mathbf{b'} = [1,3,1]$, which results in $\mathbf{q}_\text{LQ}(\mathbf{p}) = [\frac{1}{5},\frac{3}{5},\frac{1}{5}]$. 

\noindent We note that \cite{b19} uses the $L_1$, $L_2$ and $L_\infty$ norm to report worst-case distance metrics between $\mathbf{p}$ and $\mathbf{q}_{LQ}(\mathbf{p})$. By noting that the $L_1$ norm is equivalent to $2D_{TV}(\mathbf{p}, \mathbf{q}_{LQ}(\mathbf{p}))$, based on \cite{b19} the maximum source distortion between $\mathbf{p}$ and $\mathbf{q}_{LQ}(\mathbf{p})$ is as follows:\footnote{\cite{b19} proves the maximum $L_1$ distance as $\frac{1}{\ell}\frac{2a(k-a)}{k}$, where $a = \left\lfloor \frac{k}{2}\right\rfloor$. By assuming even values of $k$, the simplified expression in \eqref{eq:reznik_dis} is obtained.}: 
\begin{equation}
\label{eq:reznik_dis}
D_{TV}(\mathbf{p}, \mathbf{q}_\text{LQ}(\mathbf{p})) = \frac{k}{4\ell}.
\end{equation}
Two observations can be made from \eqref{eq:reznik_dis}. First, for high dimensional lattices, the resulting source distortion decreases, meaning that the distributions on the lattice are closer representations of $\mathbf{p}$. Second, the guarantee on the source distortion becomes looser as the length of $\mathbf{p}$ increases. This intuitively makes sense because each distribution on the lattice will have a longer length but still need to satisfy the summation constraint in \eqref{eq:lattice}, leading to distributions that are more distinct from $\mathbf{p}$. Once $\mathbf{q}_{LQ}(\mathbf{p})$ is determined, its corresponding index is calculated and transmitted. The number of bits required to send this index under this method is as follows \cite{b19}:
\begin{equation}
\label{eq:reznik_bits}
   J_{\text{LQ}} =  \left\lceil\log_2{\ell+k-1 \choose k-1}\right\rceil.
\end{equation}

\begin{figure*}
\centering
\subfloat[]{\includegraphics[scale=0.35]
{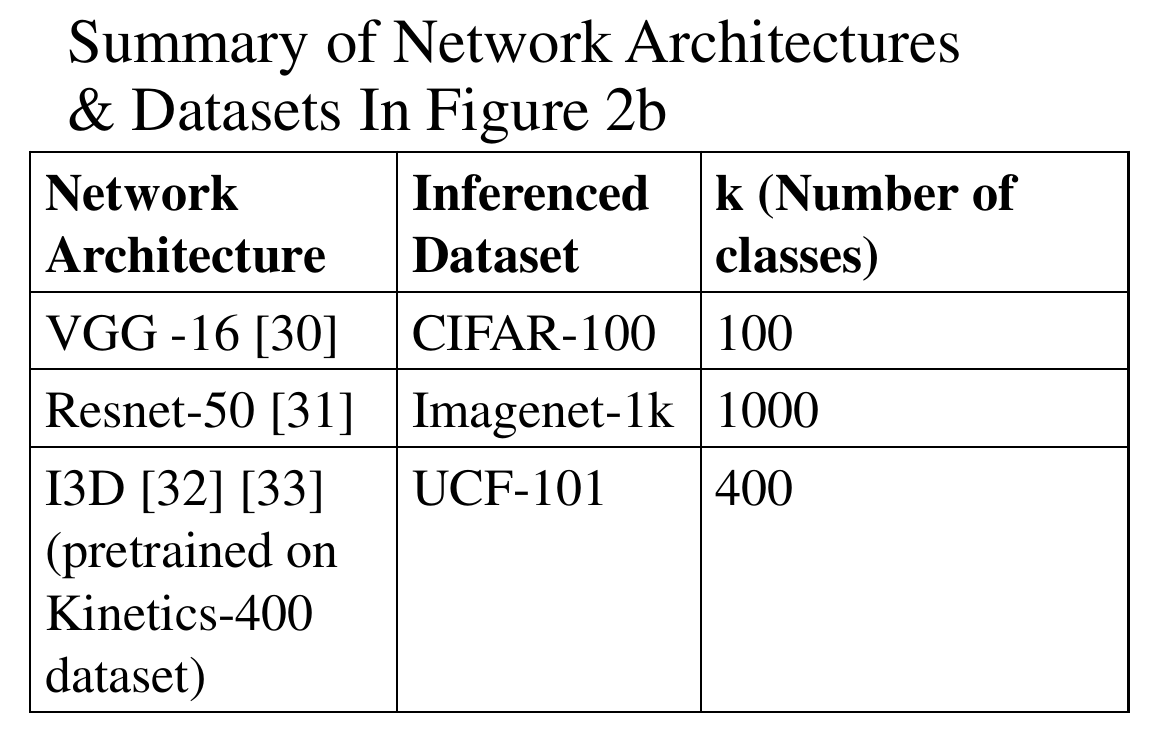}
\label{tab2}}
\subfloat[]{\includegraphics[scale=0.4]
{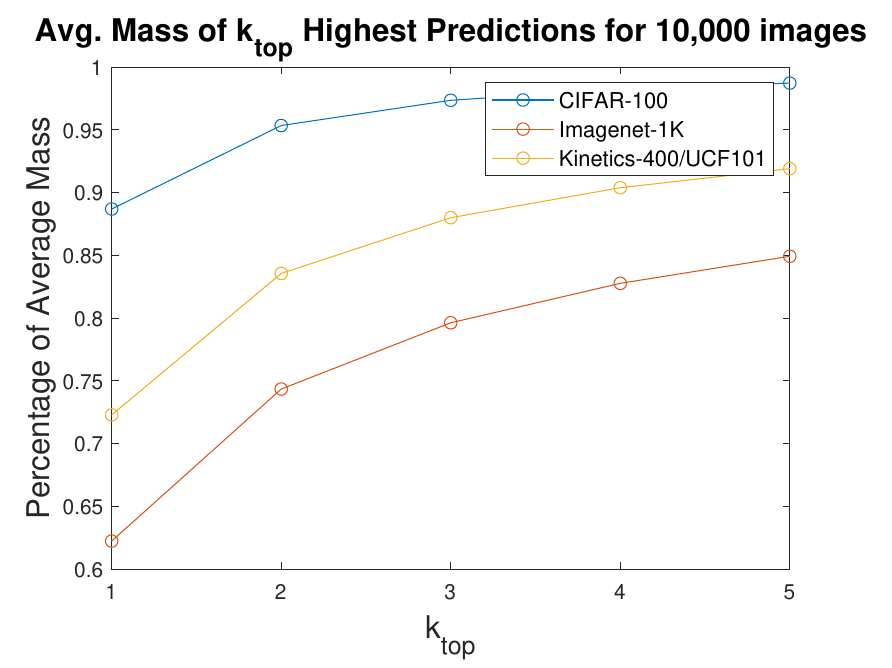}
\label{fig:cifar200second}}
\subfloat[]{\includegraphics[scale = 0.4]
{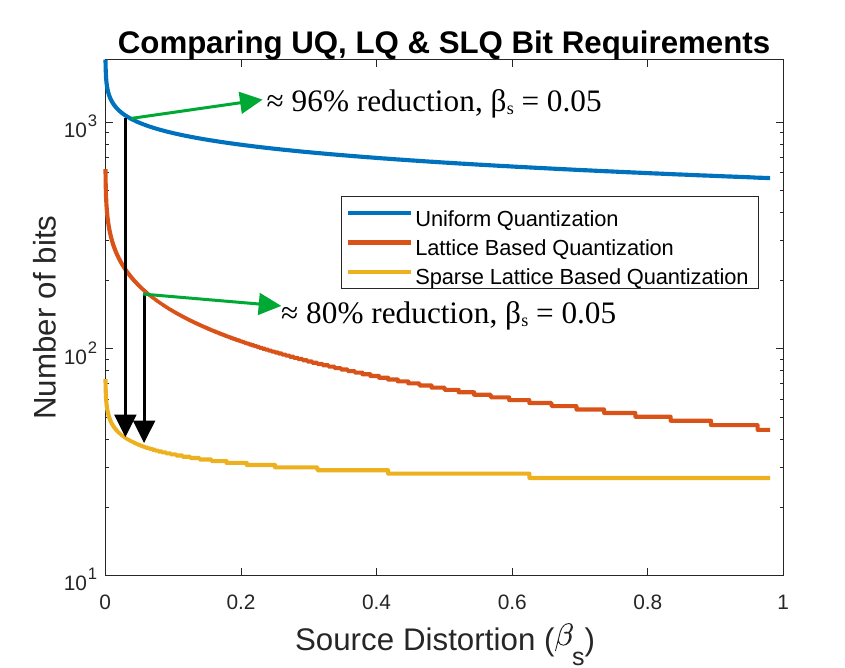}
\label{fig:bounds}}
\caption{\small (a) Table summarizing the pre-trained network architectures and datasets used to generate Figure \ref{fig:cifar200second}. (b) Percentage of average mass of $k_\text{top}$ highest predictions from different pre-trained networks and datasets across $10,000$ test images.
(c) Bit requirements for UQ, LQ, and SLQ (with $k_\text{top}= 5$ \& $\delta = 0.00001$) for different class sizes as a function of source distortion.}
\label{fig:cifar100}
\vspace{-5pt}
\end{figure*}

\noindent The next lemma shows the number of bits required under this quantization technique to attain a source distortion $\beta_s$
\begin{lemma}
\label{lma3}
For a $k$-class classification problem, if the total lattice-based quantization (LQ) budget under Algorithm \ref{alg:alg1} satisfies
\begin{align}
\label{eq:reznik}
J_{\text{LQ}} \geq  \left\lceil\log_2{\ell+k-1 \choose k-1}\right\rceil,
\end{align}
where $\ell = \left\lceil\frac{k}{4\beta_s}\right\rceil$, then $ D_\text{$TV$}(\mathbf{p},\mathbf{{q}})\leq \beta_s$.  
\end{lemma}

\noindent

\begin{remark}
    It can be observed from Lemma \ref{lma3} that the required bits for LQ has an expression similar to the required bits for UQ as shown in Lemma \ref{lma2}. However, unlike UQ, LQ does not require an additional normalization operation, which reduces the complexity of the derivation. 
\end{remark}

\noindent\subsubsection{Sparse Lattice-based Quantization (SLQ)}
For large values of $k$, it may be desired to only send a certain number of the top highest predictions in the $k$-dimensional probability vector due to many elements of the vector being very close to $0$. To accommodate this, we now propose a sparse version of the algorithm presented in \cite{b19}.  The motivation for this method comes from the notion that for high dimensional datasets (i.e. high $k$), a large portion of the mass of a classifier's decisions is concentrated in the $k_\text{top}$ highest predictions (i.e. $\sum_{i \in k_\text{top}} \mathbf{p}[i] \geq 1 - \delta$, where $0 <\delta< 1$ represents the mass of the $k-k_\text{top}$ lowest probabilities). As a case study, Figures \ref{fig:cifar200second} plots the average mass of the $k_\text{top}$ highest predictions outputted by different neural network architectures on various datasets. The networks and datasets evaluated used to generate the figure are summarized in Figure \ref{tab2}. 
Imagenet-1K \cite{b35} and CIFAR-100 \cite{b36} are variations of the Imagenet and CIFAR-10 image datasets that have been frequently used in the machine learning literature, containing 1000 and 100 classes respectively. Kinetics-400 \cite{b37} and UCF-101 \cite{b38} are datasets that have been generated for classifying human actions present in  videos, with each having 400 and 101 classes respectively.
The figure indicates that as $k_\text{top}$ increases gradually, the average mass of the probability vector encapsulated by these $k_\text{top}$ values also increases. This is beneficial as it further shows that transmitting the whole probability vector may not be required to accurately identify the classifier's decision on an image.  

Under sparse-lattice based quantization (SLQ), the $k_\text{top}$ highest values of the probability vector $\mathbf{p}$ are chosen to constitute the sparse vector $\mathbf{q}$. However, $\mathbf{q}$ does not constitute a probability vector and must be normalized, which is denoted as $\bar{\mathbf{q}}$. Algorithm \ref{alg:alg1} is then used to perform LQ on the normalized sparse vector; the resulting vector is denoted as  $\mathbf{q}_\text{SLQ}(\mathbf{p})$. The positions of the $k_\text{top}$ highest predictions also need to be transmitted for the receiver to know which classes the probabilities correspond to. We quantize the set of positions of the $k_\text{top}$ highest values by representing it as integer. The total number of bits needed to send this integer is $\left\lceil \log_2{k \choose k_\text{top}}\right\rceil$.
The total number of bits needed to send the index for $\mathbf{q}_\text{SLQ}(\mathbf{p})$ is given as
$\left\lceil\log_2{\ell+k_\text{top}-1 \choose k_\text{top}-1}\right\rceil$.
This has a similar form to the number of bits needed for the regular LQ given in \eqref{eq:reznik_bits}. It can be observed that using this procedure introduces two sources of distortion: normalization and lattice-based quantization. The next lemma shows a lower bound on the required number of bits for this technique to satisfy this more stringent restriction on the source distortion.



\noindent
\begin{lemma}
\label{lma4}
For a $k$-class classification problem, if the total sparse lattice-based quantization (SLQ) budget satisfies
\begin{align}
\label{eq:sparse_reznik}
 J_{\text{SLQ}}  \geq   \left\lceil \log_2{k \choose k_\text{top}}\right\rceil + \left\lceil\log{\ell+k_\text{top}-1 \choose k_{top}-1}\right\rceil,
\end{align}
where $\ell = \left\lceil\frac{k_\text{top}}{4(\beta_s-\delta)}\right\rceil$ \& $\sum_{i \notin k_\text{top} }\mathbf{p}[i] \leq \delta$,  then $ D_\text{$TV$}(\mathbf{p},\mathbf{q}_\text{SLQ}(\mathbf{p}))\leq \beta_s$.  

\end{lemma}

\noindent
\begin{remark}
    Despite the additional term in Lemma \ref{lma4}, compared to that of Lemma \ref{lma3}, because fewer bits are sent, the number of bits required under SLQ should be less compared to its standard counterpart for large $k$-dimensional vectors. Additionally, we also see that compared with the standard LQ, the choice of $\ell$ is also dependent on the mass encapsulated by the $k-k_\text{top}$ lowest values in the probability vector.
\end{remark}

\begin{figure*}
\centering
\subfloat[]{\includegraphics[scale=0.45]
{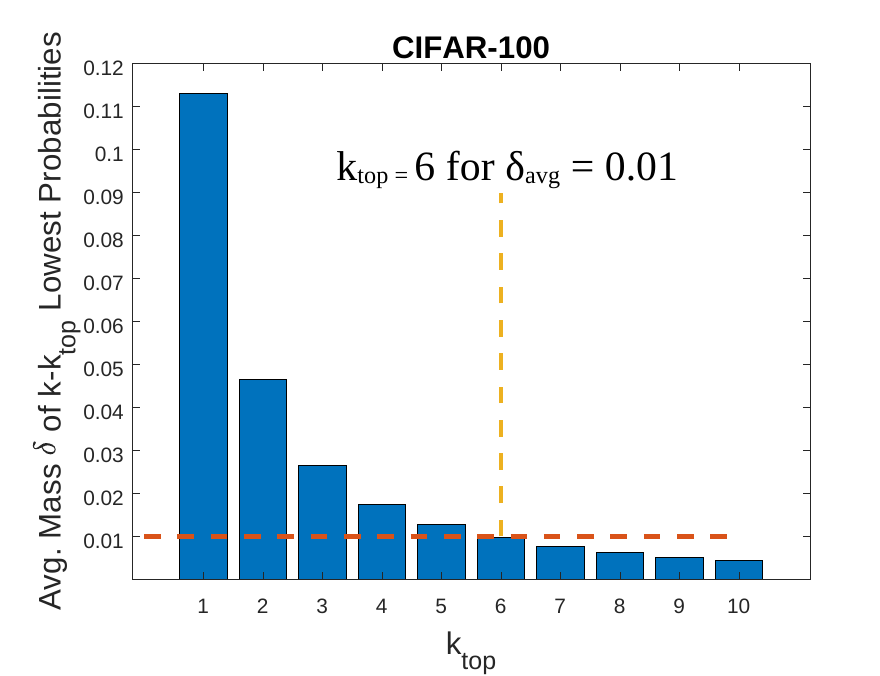}
\label{fig:cifarktop}}
\hfil
\subfloat[]{\includegraphics[scale = 0.45]
{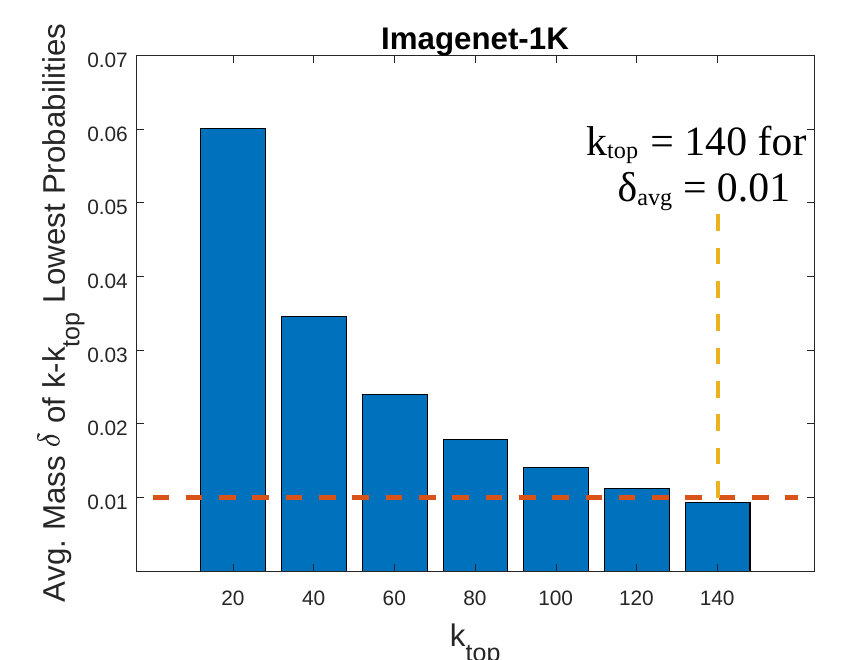}
\label{fig:imagenetktop}}
\caption{\small Average mass of $k - k_\text{top}$ lowest probabilities for different values of $k_\text{top}$ for classifications made on 10,000 images of (a) CIFAR-100 (b) Imagenet-1K.}
\label{fig:ktopselection}
\end{figure*}


\noindent \textit{Comparison of Bounds}: To develop an intuition as to how the bounds in Lemmas \ref{lma2}-\ref{lma4} compare with each other, Figure \ref{fig:bounds} plots each of them as a function of the source distortion $\beta_s$. The figure was generated assuming $k = 50$ classes and for SLQ $k_\text{top} = 5$. The figure indicates that as the allowable amount of source distortion increases, the number of bits required for UQ, LQ, and SLQ decreases. The figure also indicates that for a relatively high number of classes, SLQ requires the lowest number of bits. Looking at $\beta_s = 0.05$,  for example, SLQ incurs a reduction in bit budget of approximately $96\%$ and $80\%$ with respect to UQ and LQ. However, it can also be observed that LQ requires significantly fewer bits compared to UQ. 
Observing Lemmas 2-3 for large $k$, the number of bits for UQ and LQ an be approximated as $\mathcal{O}(k\log(\frac{k}{4\beta_s}))$ \& $\mathcal{O}(k\log(\frac{1}{4\beta_s}))$. This implies that LQ requires approximately $\mathcal{O}(\log(k))$ fewer bits compared to UQ for the same $\beta_s$. 

It is worth noting that for high $\beta_s$, the figure indicates that the number of bits required for LQ starts to approach that of SLQ. However, this phenomenon is intuitive, because when a high amount of source distortion is allowed, this implies that fewer bits are needed as the requirement to meet the total distortion constraint is  placed more on the channel encoder/decoder rather than the source encoder. Thus, we care more about the performance of the quantization schemes for low $\beta_s$, and Figure \ref{fig:bounds} indicates that SLQ significantly outperforms UQ and LQ in this regime.

\noindent \textit{Selection of $k_\text{top}$:} The latency incurred by using SLQ is contingent on the choice of $k_\text{top}$. However, $k_\text{top}$ cannot be chosen arbitrarily because the choice of $k_\text{top}$ affects the average mass of the probability vector represented in the transmitted quantized vector. Figure \ref{fig:ktopselection} shows the average mass of the $k-k_\text{top}$ lowest probabilities (denoted as $\delta_\text{avg}$) for predictions made on CIFAR-100 and Imagenet-1K. As $k_\text{top}$ increases, and more of the mass of the probability vector is subsequently represented by the $k_\text{top}$ highest values, we see that the average mass of the $k-k_\text{top}$ values decreases for both datasets. Because of the higher number of classes in Imagenet-1K, a larger choice of $k_\text{top}$ is needed to reduce $\delta_\text{avg}$ compared to CIFAR-100. If a requirement was given for $\delta_\text{avg} < 0.01$, for example, then for CIFAR-100 and Imagenet-1K this would require approximately a $k_\text{top}$ of $6$ \& $140$ respectively.

\vspace{-10pt}
\subsection{Tradeoff Between Latency \& End-to-End Distortion}

\noindent \textit{Impact of Channel Noise \& Decoding Error:}  We now incorporate the effect of channel noise and decoding errors in the analysis of the end-to-end distortion. The next lemma shows a bound on the overall expected distortion (expectation is over the channel noise realizations) if the source distortion is bounded by $\beta_s$, and the decoding error probability is given by $\epsilon^{*}(n)$.

\noindent
\begin{lemma}
\label{lma5}
For a given source distortion $\beta_s$ and decoding error probability $\epsilon^{*}(n)$, the overall expected distortion is upper bounded as follows:
\begin{equation}
     \mathbb{E}[D_\text{TV}(\mathbf{p},\hat{\mathbf{p}}(\kappa(\mathbf{y})))] \leq (1-\epsilon^{*}(n))\beta_s + \epsilon^{*}(n).
\end{equation}
\end{lemma}

\noindent
\begin{remark}
    An observation from Lemma \ref{lma5} is that there is no \textit{explicit} dependence on the specific quantization technique. The bound on overall distortion is only dependent on the source distortion $\beta_s$ and decoding error probability introduced via $\epsilon^{*}(n)$. This indicates that this framework can work generally for different quantization techniques (for instance, one could replace uniform quantization with some other sophisticated non-uniform quantizer) and the bound will only be a function of the corresponding source distortion.
\end{remark}

We next show how the results obtained up to this point can be used to devise a framework for analyzing the tradeoff between latency and distortion. In Lemma \ref{lma5}, we showed that the overall expected distortion $ \mathbb{E}[D_\text{TV}(\mathbf{p},\hat{\mathbf{p}}(\kappa(\mathbf{y}))]$ can be upper bounded by $(1-\epsilon^{*}(n))\beta_s + \epsilon^{*}(n)$. For brevity, we refer to $ \mathbb{E}[D_\text{TV}(\mathbf{p},\hat{\mathbf{p}}(\kappa(\mathbf{y}))]$ as $\beta_t$ (i.e., the total expected distortion). To derive a relationship between the overall distortion $\beta_t$ and latency $T$, we first recall the finite blocklength result from \cite{b8}, which states that the decoding error probability $\epsilon^{*}(n)$ that can be assured for sending $J$ bits through an AWGN channel is given by \cite{b8} \cite{b39}:
\begin{equation}
\label{eq:finite}
    \epsilon^{*}(n,\gamma,J)  = Q\left(\frac{nC(\gamma)-J +\frac{1}{2}\log_2n}{\sqrt{nV(\gamma)}}\right),
\end{equation}
\noindent
where $n$ represents the blocklength, $\gamma$ represents the SNR,  $C(\gamma)$ represents the capacity defined by $\frac{1}{2}\log_2(1+\gamma)$ and $V(\gamma)$ denotes the channel dispersion defined by $ \frac{\gamma(\gamma+2)}{2(\gamma+1)^2}(\log_2(e))^2$. 

Let us now return to the problem of transmitting a probability vector $\mathbf{p}$ over an AWGN channel. Observe that the number of bits one can use to represent $\mathbf{p}$ can be chosen as a function of the source distortion $\beta_s$ (via Lemmas \ref{lma2}-\ref{lma4}, i.e., $J(\beta_s)$). However, the choice of $\beta_s$, and therefore $J(\beta_s)$ also directly impact the decoding error probability $\epsilon^{*}(n,\gamma, J(\beta_s))$ as given in \eqref{eq:finite}. Thus, the resulting overall distortion from Lemma \ref{lma5} can then be bounded by $(1-\epsilon^{*}(n,\gamma, J(\beta_s)))\beta_s+ \epsilon^{*}(n,\gamma, J(\beta_s))$. Hence, if we are given a target total expected distortion of $\beta_t$, one can then optimize $\beta_s$ to minimize latency while satisfying the total distortion budget.  This is the core idea behind our approach and is formalized in the following Theorem. 
\begin{theorem}\label{thm1}
    Given a total distortion budget $\beta_\text{t}$, for a certain quantization technique we can achieve the following latency assuming an AWGN channel:
    \begin{equation}
    \label{eq:Theorem1}
        T(\beta_t) = \min_{0 \leq \beta_s \leq \beta_t} \frac{n(\beta_s)}{2B}
    \end{equation}
where 
\begin{equation}
\label{eq:quadr}
\sqrt{n(\beta_s)} = \frac{r+\sqrt{r^{2}+4C(\gamma)J(\beta_s)}}{2C(\gamma)},
\end{equation}
\noindent
and $r = \sqrt{V(\gamma)}Q^{-1}\left(\frac{\beta_t-\beta_s}{1-\beta_s}\right)$. 
\end{theorem}

\noindent
\begin{proof}
Our objective is to minimize latency while satisfying a constraint on the overall distortion  $\beta_t$. First, the number of  bits to quantize $\mathbf{p}$ can be obtained based on the choice of quantization scheme (e.g.  $J (\beta_s) = 2k\log_2\left(\frac{k}{\beta_s}\right)$ for UQ on Lemma \ref{lma2}). We can then rearrange Lemma \ref{lma5} to solve for the desired block error probability $\epsilon^{*}(n,\gamma,J(\beta_\text{s}))= \frac{\beta_t-\beta_s}{1-\beta_s}$ in terms of $\beta_t$ \& $\beta_s$. Hence, the next step is to find the minimum number of channel uses ($n$) that can support the desired block error probability of  $\frac{\beta_t-\beta_s}{1-\beta_s}$ by using  \eqref{eq:finite}. Specifically, we wish to solve for the smallest non-negative integer $n$ satisfying:
\begin{align}
    \left(\frac{\beta_t-\beta_s}{1-\beta_s}\right)\leq Q\left(\frac{nC(\gamma)-J(\beta_\text{s})+\frac{1}{2}\log_2n}{\sqrt{nV(\gamma)}}\right).\label{eq:intermediate}
\end{align}
As the $Q(\cdot)$ function (complementary CDF of standard Gaussian) is monotonically decreasing, this means that for any $n>1$, we can bound the r.h.s. of \eqref{eq:intermediate} as: 
\begin{equation}
\label{eq:Q}
    Q\left(\frac{nC(\gamma)-J(\beta_\text{s})+\frac{1}{2}\log_2n}{\sqrt{nV(\gamma)}}\right) \leq Q\left(\frac{nC(\gamma)-J(\beta_\text{s})}{\sqrt{nV(\gamma)}}\right).
\end{equation}
Thus, we can find $n$ by instead solving for the simpler equation $\left(\frac{\beta_t-\beta_s}{1-\beta_s}\right)= Q\left(\frac{nC(\gamma)-J(\beta_\text{s})}{\sqrt{nV(\gamma)}}\right)$. Applying $Q^{-1}(\cdot)$ on both sides, we arrive at the following:
\begin{equation}
\label{eq:quad}
    nC(\gamma)-\sqrt{nV(\gamma)}Q^{-1}((\beta_t-\beta_s)/(1-\beta_s)) - J(\beta_\text{s}) = 0.
\end{equation}
This equation can be viewed as a quadratic by setting $\tilde{n} = \sqrt{n}$. Solving for $n$, we arrive at the latency expression (setting $T=n/2B$). One can then optimize the latency by minimizing over all $\beta_s \in [0,\beta_t]$, thus completing the proof of Theorem \ref{thm1}. 
\end{proof}

\subsection{Generalization to Fading Channels}
\label{sec:fading}
In this section, we now extend our framework to derive results for fading channels. To accomplish this, we leverage finite blocklength results for coherent and non-coherent fading channels \cite{b30}. The  probability of error $\epsilon^{*}(n)$ for sending $J (\beta_s)$ bits through a Rayleigh fading channel assuming access to channel state information (CSI) at the receiver is given by \cite{b30}:
\vspace{-3pt}
\begin{equation}
\label{eq:finite_fading_csi}
    \epsilon^{*}(n,\gamma,J(\beta_s))  = Q\left(\frac{nC_c(\gamma)-J(\beta_s)\ln(2)}{\sqrt{nFV_c(F,\gamma)}}\right),
\end{equation}
where $F$ represents the coherence interval, $C_c$ is the capacity defined as $E[\log(1+\gamma Z_1)]$ (where $Z_1$ is a sequence of variables samples from the Gamma(1,1) distribution), and $V_c(F,\gamma)$ is the channel dispersion given as $var[\log(1+\gamma Z_1)]+\frac{1}{F}-\frac{1}{F}E[\frac{1}{1+\gamma Z_1}]^2$.

Similarly, an approximation for the error probability sending $J (\beta_s)$ bits through a Rayleigh fading channel without CSI was derived in \cite{b30} for high SNRs assuming that $0 < \epsilon^*(n) < \frac{1}{2}$. The approximation is as follows:
\begin{equation}
\label{eq:finite_fading_wout_csi}
    \epsilon^{*}(n,\gamma,J(\beta_s))  = Q\left(\frac{n\underline{I}(F,\gamma)-J(\beta_s)F \ln(2) }{\sqrt{nF\tilde{U}(F)}}\right),
\end{equation}
where $\underline{I}(F,\gamma)$ can be approximated as $(F-1)\log(F\gamma)-\log\Gamma(F)-(F-1)(1+\eta)+K'_I(F,\gamma)$, $\eta$ represents Euler's constant, $\Gamma$ is the Gamma function, $\tilde{U}(F) = (F-1)^2\frac{\pi^2}{6}+(F-1)$, and $K'_I$ is a function that must be $0$ as $\gamma \rightarrow \infty$ and $F > 2$. In this work, we assume $K'_I (F,\gamma) = \frac{F}{5\gamma}$.

With respect to \eqref{eq:finite_fading_csi} \& \eqref{eq:finite_fading_wout_csi}, recall that in this work, $J$ is determined based on a given requirement on the source distortion $\beta_s$ and choice of quantization: UQ, LO, or SLQ. Theorem \ref{thm1} introduced our framework for finding the optimal $\beta_s$ to achieve the minimum latency for a specific total distortion $\beta_t$. Following similar steps as shown in the proof for Theorem \ref{thm1}, and using \eqref{eq:finite_fading_csi} \& \eqref{eq:finite_fading_wout_csi}, the following theorems can be obtained for analyzing the latency-distortion tradeoff in fading channels with/without CSI. 

\begin{theorem}\label{thm2}
    Given a total distortion budget $\beta_\text{t}$, for a certain quantization technique we can achieve the following latency assuming a Rayleigh fading channel with CSI at the receiver:
    \begin{equation}
    \label{eq:Theorem2}
        T(\beta_t) = \min_{0 \leq \beta_s \leq \beta_t} \frac{n(\beta_s)}{2B}
    \end{equation}
where 
\begin{equation}
\label{eq:quadr_fading}
\sqrt{n(\beta_s)} = \frac{r+\sqrt{r^{2}+4C_c(\gamma)J(\beta_s)\ln2}}{2C(\gamma)},
\end{equation}
\noindent
and $r = \sqrt{FV_c}Q^{-1}\left(\frac{\beta_t-\beta_s}{1-\beta_s}\right)$. 
\end{theorem}

\begin{theorem}\label{thm3}
    Given a total distortion budget $\beta_\text{t}$, for a certain quantization technique we can achieve the following latency assuming a Rayleigh fading channel at high SNRs without CSI:
    \begin{equation}
    \label{eq:Theorem3}
        T(\beta_t) = \min_{0 \leq \beta_s \leq \beta_t} \frac{n(\beta_s)}{2B}
    \end{equation}
where 
\begin{equation}
\label{eq:quadr_fading_no_csi}
\sqrt{n(\beta_s)} = \frac{r+\sqrt{r^{2}+4\underline{I}(F,\gamma)J(\beta_s)F\ln2}}{2C(\gamma)},
\end{equation}
\noindent
and $r = \sqrt{F\tilde{U}(F)}Q^{-1}\left(\frac{\beta_t-\beta_s}{1-\beta_s}\right)$, with $0 < \frac{\beta_t-\beta_s}{1-\beta_s} < \frac{1}{2}$. 
\end{theorem}

\section{Experimental Results}
\label{sec:AA}
In this section, we present results which illustrate the tradeoff between the latency and overall distortion for sending a  probability vector to a receiver over AWGN and fading channels for UQ, LQ, and SLQ.  Unless otherwise stated, we assume that $B_0 = 10$ kHz, $\delta = 0.00001$ for SLQ, and $0 < \epsilon^{*}(n) < 0.5$.

\begin{figure*}[!t]
\centering
\subfloat[]{\includegraphics[scale = 0.42]
{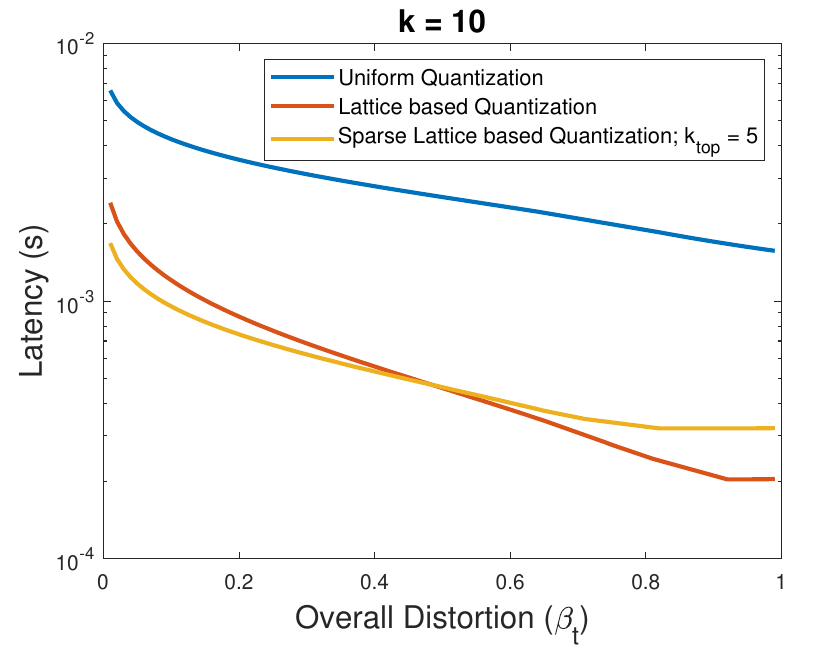}
\label{fig:4a}}
\hfil
\subfloat[]{\includegraphics[scale = 0.42]
{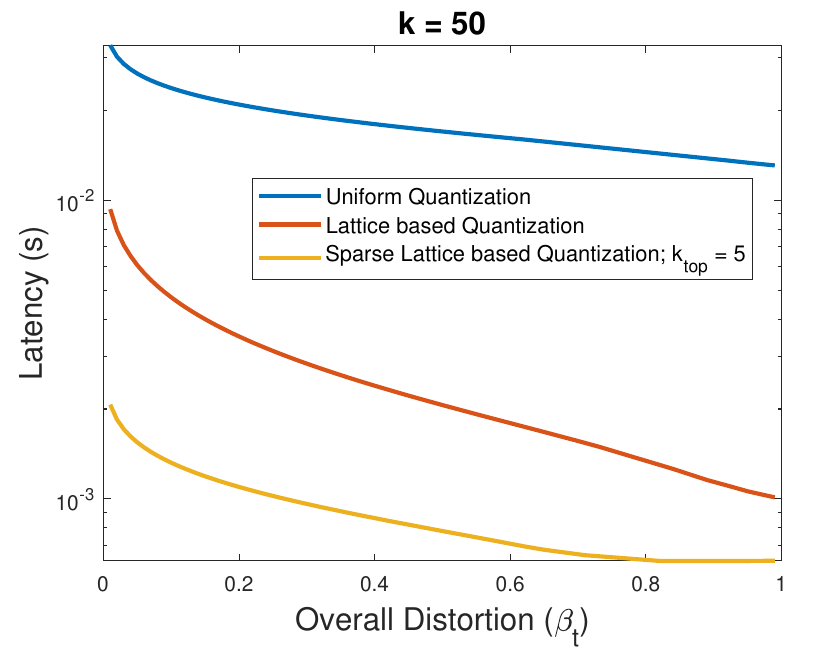}
\label{fig:4b}}
\hfil
\subfloat[]{\includegraphics[scale = 0.42]
{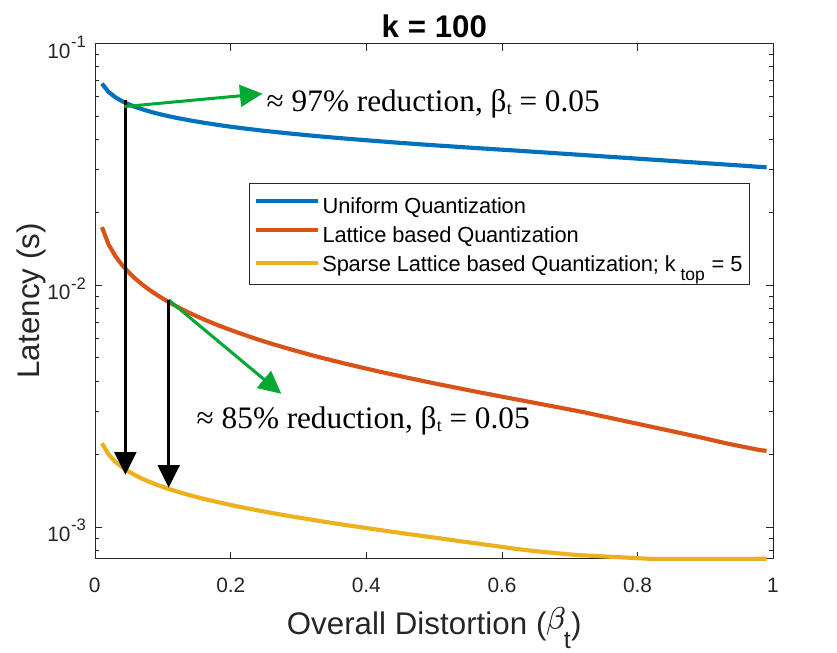} 
\label{fig:4c}}
\caption{\small Lower convex hull of latencies for different $\beta_t$ for UQ, LQ, and SLQ (obtained from Theorem \ref{thm1}). Results are reported for $k = 10 (a), 50 (b)$ \& $100 (c)$ to observe the impact varying the number of classes has on the quantization schemes.}
\label{fig:4}
\end{figure*}


\begin{figure*}[!t]
\centering
\subfloat[]{\includegraphics[scale = 0.42]
{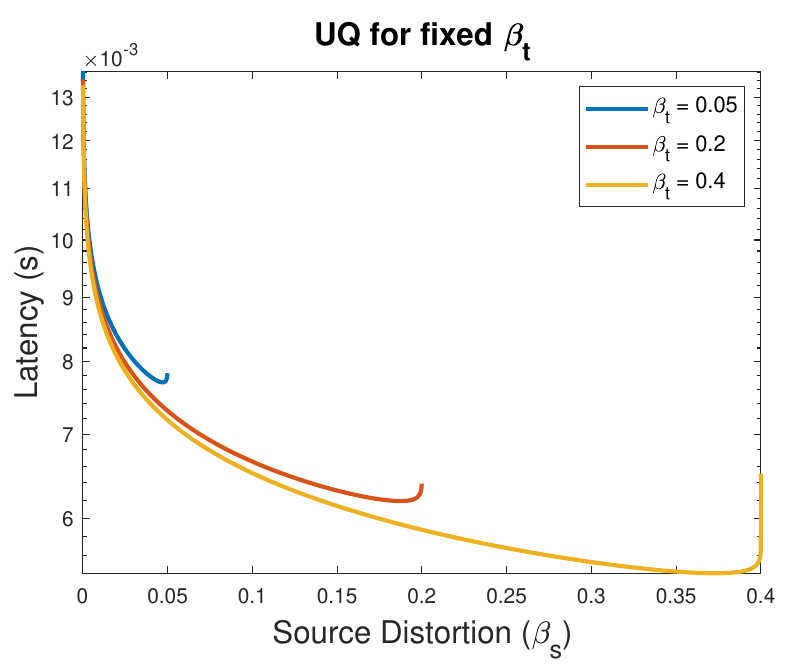}
\label{fig:5a}}
\hfil
\subfloat[]{\includegraphics[scale = 0.42]
{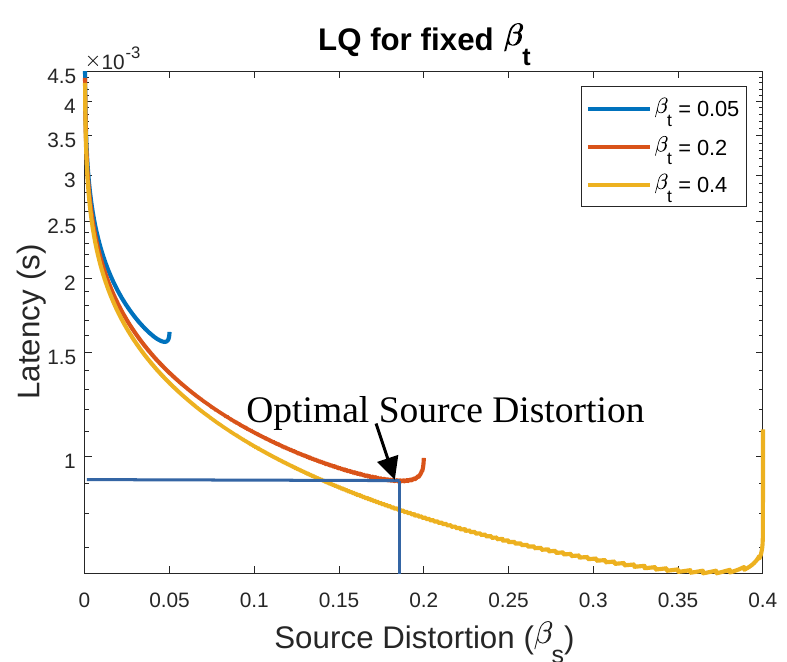}
\label{fig:5b}}
\hfil
\subfloat[]{\includegraphics[scale = 0.42]
{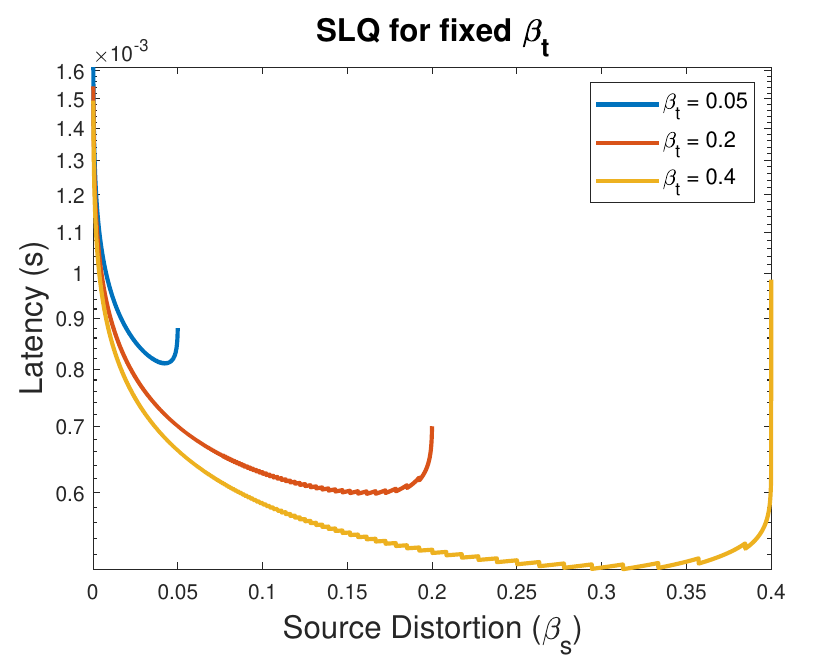}
\label{fig:5c}}
\caption{\small Impact of varying source distortion on incurring latency for fixed $\beta_t = 0.05,0.2$ \& $0.4$. Collecting results for (a) UQ, (b) LQ, (c) and SLQ assuming $k = 70$ \& $k_{top} = 20$. It can be observed that as $\beta_t$ increases, the source distortion that obtains the minimal latency also increases. It can also be observed that SLQ is able to obtain the lowest latencies out of the three techniques.}
\label{fig:5}
\end{figure*}

\subsection{Comparison of Quantization techniques}
We first observe the trade-off between latency and distortion for the uniform and lattice-based quantization techniques for the AWGN channel. 

\noindent \underline{\textit{Minimum latency for a fixed $\beta_t$ \& different k:}} Figure \ref{fig:4} reports results comparing the incurred latencies for the three quantization methods by solving the optimization problem in Theorem \ref{thm1} as the number of classes is varied for $k = 10,50$ \& $100$. We assume that $B = 320$ kHz and $\gamma_0 = 5$ dB, which yields an SNR of $\gamma \approx -10.1$ dB. We also assume $k_\text{top} = 5$ for SLQ. The figure indicates that LQ and SLQ can incur lower latencies over an AWGN channel compared to UQ as the number of classes is varied. Additionally, the figure indicates that as $k$ increases, the latency reduction from LQ to SLQ increases. At $k = 100$ and $\beta_t = 0.05$, for example, SLQ can attain a latency reduction of approximately $97\%$ and $85\%$ with respect to UQ and LQ. 
However, it's interesting to note that for low $k$ and high $\beta_t$, Figure \ref{fig:4a} indicates that LQ can incur less latencies compared to SLQ. This emphasizes that when using SLQ, more benefits are obtained at higher $k$.

\begin{figure*}
\centering
\subfloat[]{\includegraphics[scale=0.42]
{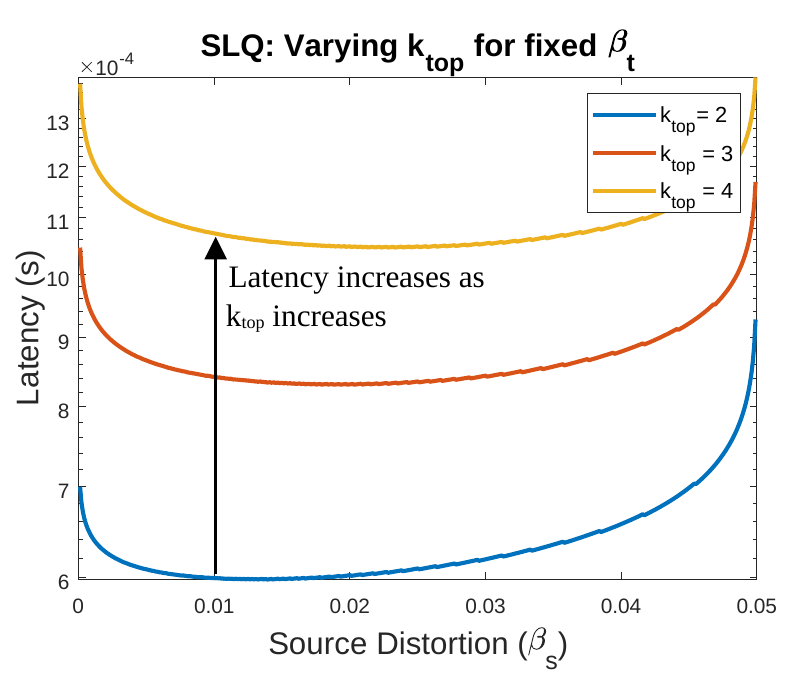}
\label{fig:diffa}}
\hfil
\subfloat[]{\includegraphics[scale = 0.42]
{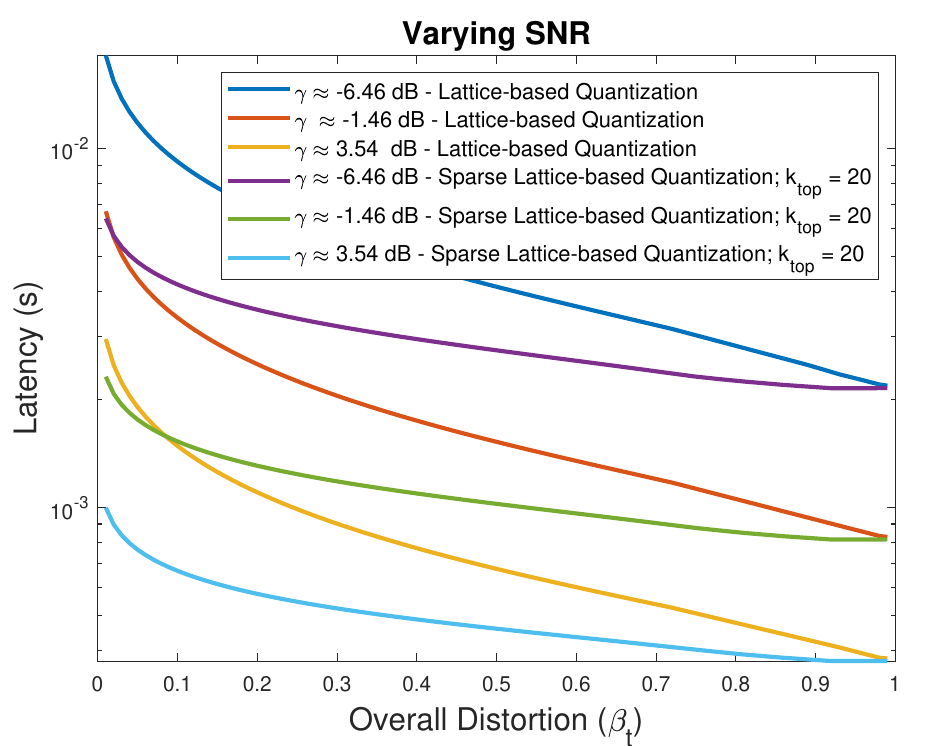}
\label{fig:diffb}}
\caption{\small (a) Impact of varying $k_{top}$ for sparse lattice-based quantization on latency as a function of source distortion, with $k = 1000$ classes and $\beta_t = 0.05$. As expected, a lower choice of $k_\text{top}$ incurs lower latency. (b) Lower convex hull of observed latencies as a function of $\beta_\text{t}$ for different SNRs ($-6.46, -1.46, 3.54$ dB) with $k = 100$ classes.}
\label{fig:diff}
\end{figure*}

\noindent \underline{\textit{Interplay between source distortion ($\beta_s$) and Latency:}} Fig. \ref{fig:5} compares the quantization schemes for a fixed total distortion $\beta_t$ with $k = 70$ classes and $k_{top} = 20$ for SLQ. We also set $B = 100$ kHz and $\gamma_0 = 15$ dB, which results in $\gamma = 5$ dB. Each of the figures indicate, that for higher dimensional probability vectors, a reduction in latency can be achieved when more source distortion is allowed. However, as the total distortion is increased, the optimal source distortion increases for each of the lattice-based methods. The figure also indicates that the lattice based quantizers can attain lower latencies compared to UQ, with SLQ performing better than LQ when quantizing high dimensional probability vectors. Each of the figures also indicate that surges in the latency occur as $\beta_s$ approaches $\beta_t$.  This intuitively makes sense because as $\beta_s$ approaches $\beta_t$, this implies that no compensation for the
distortion is being performed by the source encoder/decoder. This means that the channel encoder/decoder are responsible
for satisfying the requirement on $\beta_t$ and can only do so by
using higher blocklengths.

\noindent \underline{\textit{Impact of degree of sparseness:}} \noindent
Figure \ref{fig:diffa} presents the effect of varying $k_\text{top}$ (i.e. how many of the highest probabilities are chosen for transmission) has on the latency for a fixed $\beta_t$ when using SLQ. It's assumed that $B = 100$ kHz and $\gamma_0 = 8$ dB resulting in $\gamma = -2$ dB. Similar to Figure \ref{fig:4a}, to observe the full merit of the method, $k = 1000$. The figure indicates that as $k_\text{top}$ increases, the latency also increases, which is as expected as this means more predictions are included in the sparse vector. The figure also indicates, similar to Figure \ref{fig:5}, that as more source distortion is allowed, smaller latencies can be attained. This means that to quantize and send additional values from the probability vector at a lower latency, more source distortion must be allowed.

\noindent \underline{\textit{Latency as a function of SNR:}} Figure \ref{fig:diffb} presents the latencies incurred for LQ and SLQ at different SNRs. Recall that the SNR is related to the bandwidth $B$ as $\gamma = \frac{\gamma_0B_0}{B}$; thus by varying the reference SNR $\gamma_0$, different SNRs $\gamma$ can be simulated. In Figure \ref{fig:diffb}, $\gamma_0$ is varied to $5$ dB, $10$ dB, and $15$ dB respectively, which corresponds to $\gamma$ of $-6.46, -1.46, 3.54$ dB respectively. To observe the full benefits of SLQ, $k = 100$ and $k_\text{top} = 20$. 
The figure indicates that for both techniques, the incurred latency decreases as the SNR increases. Figure \ref{fig:diffb} also indicates that SLQ significantly outperforms LQ at each of the simulated SNRs. This means that for a high $k$-dimensional probability vector, SLQ can incur lower latencies even in poor channel conditions.

\subsection{Application to Fading Channels}
\begin{figure*}[!t]
\centering
\subfloat[]{\includegraphics[scale=0.45]
{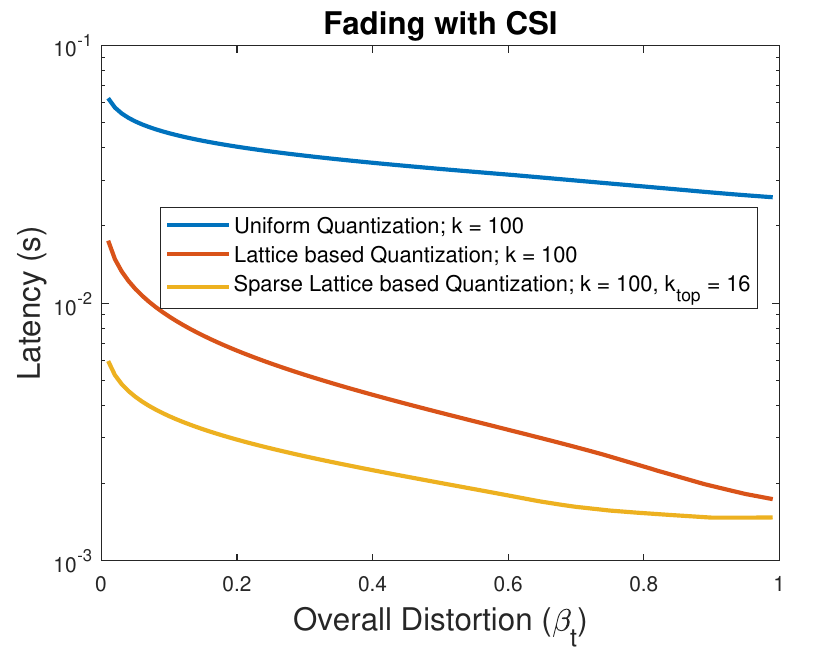}
\label{fig:6a}}
\hfil
\subfloat[]{\includegraphics[scale=0.45]
{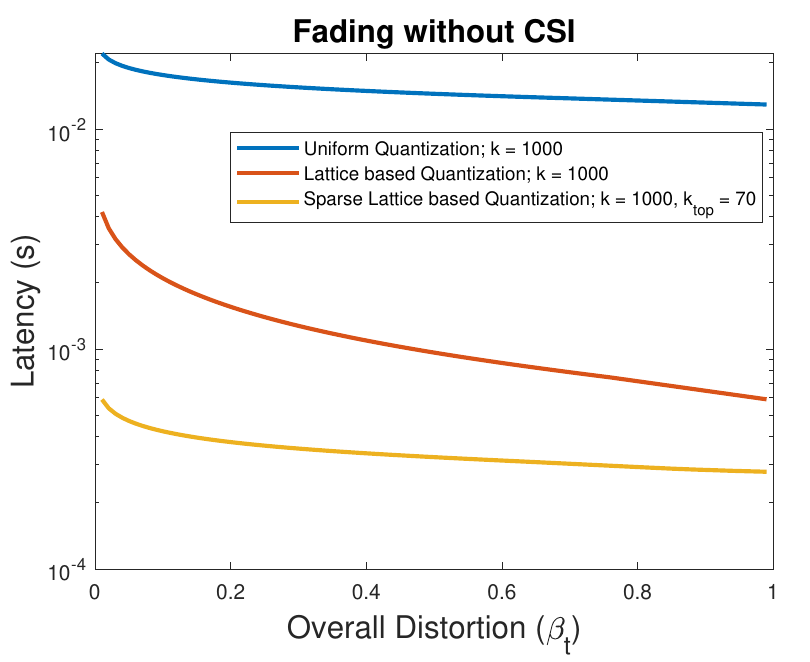}
\label{fig:6b}}
\caption{\small(a) Lower convex hull of latencies for different $\beta_t$ for UQ, LQ, and SLQ for a Rayleigh fading channel assuming CSI at the receiver. (b) Lower convex hull of latencies for different $\beta_t$ for UQ, LQ, and SLQ for a Rayleigh fading channel without CSI at the receiver.
}
\label{fig:6}
\end{figure*}

We now analyze our framework considering Rayleigh block-fading channels using the finite blocklength approximations presented in Section \ref{sec:fading}. We assume $k_{top} = 16$, $B = 100$ kHz, $\gamma_0 = 11$ dB, and $k = 100$, which results in $\gamma = 1$ dB. Figure \ref{fig:6a} shows the latency-distance tradeoff for the three quantization schemes for a Rayleigh fading channel assuming CSI at the receiver and $F = 20$. The figure indicates that, similar to our previously presented results, SLQ can still outperform UQ and LQ. Figure \ref{fig:6b}  shows the latency-distortion tradeoff for the three quantization schemes for a Rayleigh fading channel without CSI. As \eqref{eq:finite_fading_wout_csi} is a high SNR approximation, we have adjusted the following parameters: $B = 200 kHz$, $B_0 = 800 kHz$, $\gamma_0 = 15$ dB, which results in $\gamma \approx 21$ dB. Additionally, $k$ is set to 1000 classes, with $k_\text{top} = 70$ for SLQ. It can be observed that even without access to CSI, similar trends are observed with SLQ performing significantly better than UQ \& LQ.


\section{Conclusion}
\label{sec:conc}
\vspace{-2pt}
In this work, we have investigated a framework where the decisions (a probability vector) from a classification task are transmitted over a noisy channel. Specifically, we study the tradeoff between the latency associated with transmitting this result against the distortion incurred with quantizing the result and the impact of channel noise on the transmission. To accomplish this, we have analyzed the performance of uniform and lattice-based quantization techniques by first providing results bounding the necessary bit budgets under each technique to satisfy a requirement on the allowable source distortion. Then by linking distortion due to decoding errors (using results from finite blocklength channel capacity) with the distortion due to quantization, we are able to create a framework that allows us to find an optimized source distortion that achieves a minimal transmission latency at different levels of end-to-end distortion. Our results show that there is an interesting interplay between source distortion (i.e., distortion for the probability vector measured via f-divergence) and the subsequent channel encoding/decoding parameters; and indicate that a \textit{joint} design of these parameters is crucial to navigate the latency-distortion tradeoff. After varying different parameters of the framework, and assuming both AWGN and fading channels, our results show that sparse-lattice based quantization is the most efficient at minimizing latency at different levels of end-to-end distortion. Specifically, our results indicate that sparse-lattice based quantization outperforms all other methods for high dimensional probability vectors (i.e. a higher number of classes) and sparse predictions generated by the classifier (which is often the case in various ML classifiers, as also evidenced in CIFAR-100, Imagenet-1K, and Kinetics-400 datasets).  We believe that the sparse lattice based quantization techniques could also be useful for other ML based systems requiring low latency, such as in transmitting semantic information. 

{\appendix[Proof of Lemma \ref{lma1}]
The minimal achievable latency $T^{*}(\beta)$ is a non-increasing function of $\beta$. This is clear from the fact that any decoder which satisfies a distortion constraint of $\beta$ also satisfies the distortion constraint of $\beta'$ for $\beta'\geq \beta$. 

We next show that $T^{*}(\beta)$ is a convex function of $\beta$. 
Let $T^{*}(\beta_{t_1})$  ($T^{*}(\beta_{t_2})$, respectively) represent the minimum latencies obtained using encoder-decoder pair $(\mathcal{E}_\text{1}^{*},\mathcal{D}_\text{1}^{*})$ ($(\mathcal{E}_\text{2}^{*},\mathcal{D}_\text{2}^{*})$, respectively) that satisfy $D_\text{f}(\mathbf{p},\hat{\mathbf{p}_\text{1}}) \leq \beta_{t_1}$ ($D_\text{f}(\mathbf{p},\hat{\mathbf{p}}_\text{2}) \leq \beta_{t_2}$, respectively). 
We define a new encoder-decoder pair $(\mathcal{E}_\text{3},\mathcal{D}_\text{3})$ such that,
\[
(\mathcal{E}_\text{3},\mathcal{D}_\text{3}) =
\begin{cases} 
      \label{combinations}(\mathcal{E}_\text{1}^{*},\mathcal{D}_\text{1}^{*}) & \text{with probability } \alpha \\
      (\mathcal{E}_\text{2}^{*},\mathcal{D}_\text{2}^{*}) & \text{with probability }  1-\alpha. \\
     
   \end{cases}
\]

The expected latency using $(\mathcal{E}_\text{3},\mathcal{D}_\text{3})$ is $\alpha T^{*}(\beta_{t_1}) + (1-\alpha) T^{*}(\beta_{t_2})$. The total distortion using $(\mathcal{E}_\text{3},\mathcal{D}_\text{3})$ is $D_\text{f}(\mathbf{p},\alpha\hat{\mathbf{p}}_\text{1}+ (1-\alpha)\hat{\mathbf{p}}_\text{2})$, which can be upper bounded as, 
\begin{equation}
\begin{aligned}
    D_\text{f}(\mathbf{p},\alpha\hat{\mathbf{p_\text{1}}}+ (1-\alpha)\hat{\mathbf{p}}_\text{2}) & \stackrel{(a)}{\leq} \alpha D_\text{f}(\mathbf{p},\hat{\mathbf{p}}_\text{1}) + (1-\alpha) D_\text{f}(\mathbf{p},\hat{\mathbf{p}}_\text{2})\\
    & \stackrel{(b)}{\leq} \alpha \beta_{t_1} + (1-\alpha) \beta_{t_2},
\end{aligned}
\end{equation}
where (a) follows from convexity of $f$-divergence, and (b) follows from the bounds on end-to-end distortion for the two individual decoders. Let us now consider the optimal encoder-decoder pair $(\mathcal{E}^{*},\mathcal{D}^{*})$ that satisfies the distortion constraint $D_\text{f}(\mathbf{p},\hat{\mathbf{p}}) \leq \alpha\beta_{t_1} + (1-\alpha)\beta_{t_2}$. The minimum latency using $(\mathcal{E}^{*},\mathcal{D}^{*})$ is then $T^{*}(\alpha\beta_{t_1} + (1-\alpha)\beta_{t_2})$. Recall, $(\mathcal{E}^{*},\mathcal{D}^{*})$ is optimal encoder-decoder pair, and therefore, the corresponding latency must be always less than the latencies obtained using any $(\mathcal{E}_\text{3},\mathcal{D}_\text{3})$ pair. That is,
\begin{align}
T^{*}(\alpha \beta_{t_1} + (1-\alpha) \beta_{t_2}) \leq \alpha T^{*}(\beta_{t_1}) + (1-\alpha) T^{*}(\beta_{t_2}).\nonumber 
\end{align}
This proves that $T^{*}(\beta)$ is convex. $D^{*}(T)$ can be shown to be convex in a similar manner by leveraging the convexity of $f$-divergence.  
\vspace{-\baselineskip}
\section*{Proof of Lemma \ref{lma2}}
We consider uniform quantization (UQ) with bins of width $\frac{1}{2^j}$, where $j= \lfloor{J/k}\rfloor$. For $r \in \{1,2,\cdots,2^{j}-1\}$, we define the values for the $r^{\text{th}}$ bin in the range $\left[\frac{r}{2^j}, \frac{r+1}{2^j}\right)$. In other words, for any $i \in [k]$, $\mathbf{q}[i]$ is obtained by mapping the value of $\mathbf{p}[i]$ in the range $\left[\frac{r_i}{2^j}, \frac{r_i+1}{2^j}\right)$ to $\frac{r_i+\frac{1}{2}}{2^j}$. However, we note that $\mathbf{q}$ may not necessarily be a probability vector. We define the probability vector $\mathbf{\tilde{q}}$, by normalizing the values in  $\mathbf{q}$.  Therefore, for any $i \in [k]$, we can write $\mathbf{p}[i]$ and $\mathbf{\tilde{q}}[i]$ as follows: 
\begin{align}
    \mathbf{p}[i] = \frac{r_i+\delta_i}{2^j}, \hspace{1cm} \mathbf{\tilde{q}}[i] = \frac{r_i+\frac{1}{2}}{S \cdot 2^j },
\end{align}
where $\delta_i \in [0,1)$ and $S = {\sum_{i=1}^{k}}\frac{r_i+\frac{1}{2}}{2^j}$. Returning to our goal, recall that we wish to pick $J$ such that $D_\text{TV}(\mathbf{p}, \mathbf{\tilde{q}}) \leq \beta_s$. To this end, we first bound  
\begin{align}
    \label{eq:lma2_step2}
        D_\text{TV}(\mathbf{p}, \mathbf{\tilde{q}}) & = \frac{1}{2}\sum_{i=1}^{k}|\mathbf{p}[i] - \mathbf{\tilde{q}}[i]| \nonumber \\ 
        & = \frac{1}{2}\sum_{i=1}^{k}\left| \frac{r_i+\delta_i}{2^j} - \frac{r_i+\frac{1}{2}}{S \cdot 2^j}\right| \nonumber \\  & = 
        \frac{1}{2^{(j+1)}}\sum_{i=1}^{k}\left| r_i \left( 1-\frac{1}{S} \right)
        +\delta_i - \frac{1}{2S}\right| \nonumber \\ & \stackrel{(a)}{\leq}
        \frac{1}{2^{(j+1)}}\sum_{i=1}^{k} \left(\left|r_i \left(1-\frac{1}{S} \right)\right| + \left|\delta_i - \frac{1}{2S}\right|\right) \nonumber \\ & \stackrel{(b)}{\leq}
        \frac{1}{2^{(j+1)}}\sum_{i=1}^{k}\left(\left|2^{j} \left(1-\frac{1}{S}\right)\right| + \left|\delta_i - \frac{1}{2S}\right|\right),
\end{align}
where (a) follows from triangle inequality, and (b) follows from the fact that $r_i < 2^j$. We also know that, $\sum_{i}^{k} \mathbf{p}[i] = 1$ and $\sum_{i}^{k} \mathbf{\tilde{q}}[i] = S$. Consider the difference, 

\begin{equation}
\label{eq:lma2_step1}
\left|\sum_{i=1}^{k} \frac{r_i+\delta_i}{2^j} - \sum_{i=1}^{k}\frac{r_i+\frac{1}{2}}{2^j}\right| \leq \sum_{i=1}^{k} \frac{|\delta_i-\frac{1}{2}|}{2^j} \leq \frac{k}{2^{j+1}}.
\end{equation}
Therefore, we have that, 
\begin{equation}
\label{eq:sbounds1}
    |1-S| \leq \frac{k}{2^{j+1}}.
\end{equation}
Suppose that $j$ is given as 
\begin{align}
\label{eq:assumption}
j = \log_2\left(\frac{k}{2\alpha}\right) 
\end{align}
for some $\alpha \in (0,0.5]$.  Therefore, from \eqref{eq:sbounds1} we have that, 
\begin{equation}
\label{eq:sbounds2}
    1-\alpha \leq S \leq 1+\alpha.
\end{equation}
Using \eqref{eq:assumption} and \eqref{eq:sbounds2} in \eqref{eq:lma2_step2}, we can further bound $D_\text{TV}(\mathbf{p}, \mathbf{\tilde{q}})$ as,
\begin{align}
\label{eq:lma2_beforebeta}
    D_\text{TV}(\mathbf{p},\mathbf{\tilde{q}}) & \leq  \left[\frac{\alpha k}{2(1-\alpha)}+\max \left\{ {\frac{\alpha}{2(1-\alpha)}, \alpha - \frac{\alpha}{2(1+\alpha)}} \right\} \right] \nonumber \\ 
    & \stackrel{(a)}{\leq} \frac{\alpha k}{2(1-\alpha)} + \frac{\alpha}{2(1-\alpha)} = \frac{\alpha}{1-\alpha}\left(\frac{k+1}{2}\right),
\end{align}
where (a) holds for all $\alpha \in (0,0.5]$. Now, since we require $D_\text{TV}(\mathbf{p},\mathbf{\tilde{q}}) \leq  \beta_s$, we can pick $\alpha$ such that $\frac{\alpha}{1-\alpha}\left(\frac{k+1}{2}\right)\leq \beta_s$. We can pick  $\alpha$ to satisfy this constraint with equality, i.e., 
\begin{align}
\label{eq:final-alpha}
\alpha^{*} = \frac{2\beta_\text{s}}{k+1+\beta_\text{s}}.
\end{align}
Next, substituting \eqref{eq:final-alpha} in \eqref{eq:assumption},  we can then claim that as long as the total bit budget  $J =k j \geq k\log_2(k/2\alpha^*)$, then $D_\text{TV}(\mathbf{p},\mathbf{\tilde{q}}) \leq  \beta_s$. Now, using the fact that $k \geq 2$  and $\beta_s \in [0,1]$, it can be readily verified that $2k \cdot \log_2 \left(\frac{k}{\beta_s}\right) \geq k\log_2(k/2\alpha^*)$,   completing the proof of Lemma \ref{lma2}.

\section*{Proof of Lemma \ref{lma3}}
We denote the probability distribution on $Q_\ell$ closest to a given probability vector $\mathbf{p}$ as $\mathbf{q}_\text{LQ}(\mathbf{p})$. Recall from \cite{b19} that by performing lattice-based quantization (LQ),  the distortion incurred is given by $D(\mathbf{p},\mathbf{q}_\text{LQ}(\mathbf{p})) = \frac{k}{4\ell}$. For our framework, we propose setting $\ell = \left\lceil\frac{k}{4\beta_s}\right\rceil$. In doing so, we can obtain the following upper bound on $D(\mathbf{p},\mathbf{q}_\text{LQ}(\mathbf{p}))$:
\begin{align}
    \label{eq:lma2_step22}
        D(\mathbf{p},\mathbf{q}_\text{LQ}(\mathbf{p}))& =
        \frac{k}{4\ell} \nonumber \\ 
        & = \frac{k}{4\left\lceil\frac{k}{4\beta_s}\right\rceil}\nonumber \\  & \leq  \frac{k}{4\left(\frac{k}{4\beta_s}\right)} 
         \nonumber \\ & = \beta_s.  \nonumber
\end{align}
 Thus, sending $\left\lceil\log_2{\ell+k-1 \choose k-1}\right\rceil$ bits under LQ, with $\ell = \left\lceil\frac{k}{4\beta_s}\right\rceil$ will satisfy the source distortion requirement. 
 
\section*{Proof of Lemma \ref{lma4}}
For sparse lattice-based quantization (SLQ), we assume $\mathbf{p}$ has the following property: $\sum_{i \in k_\text{top}} \mathbf{p}[i]\geq 1 - \delta$, where $\delta \in [0,1]$. In other words, we assume that the $k_\text{top}$ highest values constitute a significant portion of the mass of $\mathbf{p}$. This implies that $\sum_{i \notin k_\text{top}} \mathbf{p}[i]\leq \delta$. Let $S = \sum_{i \in k_\text{top}} \mathbf{p}[i]$. We denote $\bar{\mathbf{q}}$ as the resulting probability vector normalized by the sum of the $k_{top}$ values; more explicitly, $\bar{\mathbf{q}}[i] =\frac{\mathbf{p}[i]}{S}$ if $ i \in k_\text{top}$ and zero otherwise. Lastly, $\mathbf{q}_\text{SLQ}(\mathbf{p})$ is the subsequent probability vector after passing the non-zero values of $\bar{\mathbf{q}}$ into the standard LQ algorithm (Algorithm \ref{alg:alg1}).

\noindent To determine the number of bits needed to send the index of $\mathbf{q}_\text{SLQ}(\mathbf{p})$ we want to bound the bit budget as a function of the source distortion incurred. Under SLQ, there are two causes of distortion: normalization of the $k_\text{top}$ highest values and standard LQ. To represent this, we first prove the following statement on the distortion encapsulated by both operations:
\begin{equation}
    \label{eq: endtoendidst}
   D_\text{$TV$}(\mathbf{p}, \mathbf{q}_\text{SLQ}(\mathbf{p})) \leq D_\text{$TV$}(\mathbf{p}, \bar{\mathbf{{q}}}) + D_{TV}(\bar{\mathbf{q}}, \mathbf{q}_\text{SLQ}(\mathbf{p})),
\end{equation}
\noindent  where $D_{TV}(\mathbf{p},\bar{\mathbf{q}})$ represents the distortion incurred through normalization and $D_{TV}(\bar{\mathbf{q}}, \mathbf{q}_\text{SLQ}(\mathbf{p}))$ represents the distortion incurred through LQ. We prove \eqref{eq: endtoendidst} as follows:
\begin{align}
    \label{eq:triangle}
        D_{TV}(\mathbf{p},\mathbf{q}_\text{SLQ}(\mathbf{p})) & = \frac{1}{2} |\mathbf{p} - \mathbf{q}_\text{SLQ}(\mathbf{p})|\nonumber \\ 
        &= \frac{1}{2} |\mathbf{p} - \bar{\mathbf{q}} + \bar{\mathbf{q}} - \mathbf{q}_\text{SLQ}(\mathbf{p})|\nonumber \\  & \stackrel{(a)}{\leq} \frac{1}{2} |\mathbf{p} - \bar{\mathbf{q}}| + \frac{1}{2} |\bar{\mathbf{q}} - \mathbf{q}_\text{SLQ}(\mathbf{p})|
         \nonumber \\ & = D_{TV}(\mathbf{p},\bar{\mathbf{q}}) + D_{TV}(\bar{\mathbf{q}},\mathbf{q}_\text{SLQ}(\mathbf{p})), \nonumber
\end{align}
\noindent where (a) follows from the triangle inequality.  Having proved \eqref{eq: endtoendidst}, we now upper-bound $ D_\text{TV}(\mathbf{p}, \mathbf{q}_\text{SLQ}(\mathbf{p}))$ by the required source distortion $\beta_s$, which can be explicitly stated as: 
\begin{equation}
\label{eq:sparsesource}
  D_\text{$TV$}(\mathbf{p}, \mathbf{q}_\text{SLQ}(\mathbf{p})) \leq D_\text{$TV$}(\mathbf{p}, \bar{\mathbf{q}}) + D_{TV}(\bar{\mathbf{q}}, \mathbf{q}_\text{SLQ}(\mathbf{p})).
\end{equation}

\noindent We upper bound $D_{TV}(\mathbf{p},\bar{\mathbf{q}})$ as follows:
\begin{align}
            D_\text{$TV$}(\mathbf{p}, \bar{\mathbf{q}}) & = \frac{1}{2}\sum_{i}|\mathbf{p}[i] - \bar{\mathbf{q}}[i]| \nonumber \\ 
        & = \frac{1}{2}\left(\sum_{i \in k_\text{top}}\left| \mathbf{p}[i] - \frac{\mathbf{p}[i]}{S}\right| \nonumber +  \sum_{i \notin k_{top}} |\mathbf{p}[i]- \mathbf{\bar{q}}[i]|\right) \\  & \stackrel{(a)}= 
        \frac{1}{2}\left(\sum_{i \in k_\text{top}}\mathbf{p}[i]\left| 1-\frac{1}{S}
\right| + \sum_{i \notin k_\text{top}} |\mathbf{p}[i]|\right)\nonumber \\ &=
         \frac{1}{2}\left(\sum_{i \in k_\text{top}} \mathbf{p}[i]\left|\frac{S-1}{S} \right| + \sum_{i \notin k_\text{top}} |\mathbf{p}[i]|\right)\nonumber \\ & \stackrel{(b)}=  \frac{1}{2}\left(\sum_{i \in k_\text{top}} \mathbf{p}[i]\frac{|S-1|}{S} + \sum_{i \notin k_\text{top}} \mathbf{p}[i] \right) 
         \end{align}
where (a) follows from $\bar{\mathbf{q}}[i] = 0 \ \forall i \notin k_\text{top}$, (b) follows from $S \geq 0$.
         \begin{align}
          D_\text{$TV$}(\mathbf{p}, \bar{\mathbf{q}}) &\stackrel{(a)}=  \frac{1}{2}\left(\sum_{i \in k_\text{top}} \mathbf{p}[i]\frac{1-S}{S} + \sum_{i \notin k_\text{top}} \mathbf{p}[i] \right)\nonumber \\  & =  \frac{1}{2}\left(\sum_{i \in k_{top}} \frac{\mathbf{p}[i]}{S} -\sum_{i \in k_\text{top}} \mathbf{p}[i] + \sum_{i \notin k_{top}} \mathbf{p}[i] \right)
\nonumber \\   & \stackrel{(b)}{=}
         \frac{1}{2}\left(1 - \sum_{i \in k_\text{top}} \mathbf{p}[i] + \sum_{i \notin k_\text{top}} \mathbf{p}[i] \right)\nonumber \\
         & \stackrel{(c)}{=}
        \left(1 - \sum_{i \in k_\text{top}} \mathbf{p}[i]\right)\nonumber \\
        & \stackrel{(d)}{\leq}
        \delta, \nonumber \\
\end{align}

 where (a) follows from $ 0 \leq S \leq 1$, (b) follows from $S = \sum_{i \in k_\text{top}} \mathbf{p}[i]$, (c) follows from $\sum_{i \notin k_\text{top}} \mathbf{p}[i]  = 1 - \sum_{i \in k_\text{top}} \mathbf{p}[i]$, (d) follows from $\sum_{i \notin k_\text{top}} \mathbf{p}[i]  = 1 - \sum_{i \in k_\text{top}} \mathbf{p}[i] \leq \delta$. 

\noindent From \eqref{eq:reznik_dis}, we can upper bound the distortion due to LQ as $D_{TV}(\bar{\mathbf{q}}, \mathbf{q}_\text{SLQ}(\mathbf{p})) \leq \frac{k_\text{top}}{4\ell}$, as only the $k_\text{top}$ non-zero values of $\bar{\mathbf{q}}$ will be passed into Algorithm \ref{alg:alg1} for quantization. Recalling \eqref{eq:reznik_bits}, this would imply that  $\left\lceil\log_2{\ell+k_{top}-1 \choose k_{top}-1}\right\rceil$ bits are needed to send the index of the resulting probability vector. Substituting these bounds into \eqref{eq:sparsesource} gives
\begin{equation}
\label{eq:8}
   \delta + \frac{k_\text{top}}{4\ell} \leq \beta_s.
\end{equation}

\noindent Solving for $\ell$ in $\eqref{eq:8}$ gives: 
\begin{equation}
\label{eq:bound_on_l}
   \ell  =  \left\lceil\frac{k_\text{top}}{4(\beta_s-\delta)}\right\rceil,
\end{equation}
\noindent which implies that $\beta_s > \delta$. We now have a bound on the choice of $\ell$ to ensure a source distortion no greater than $\beta_s$ accounting for normalization and standard LQ. The positions of the $k_\text{top}$ highest predictions also need to be transmitted for the receiver to know which classes the probabilities correspond to. To address this, the set of positions of the $k_\text{top}$ highest values are represented as an integer. This means that $\left\lceil \log_2{k \choose k_\text{top}}\right\rceil$ bits are required to send this integer. Thus, $\left\lceil\log_2{\ell+k_{top}-1 \choose k_{top}-1}\right\rceil + \left\lceil \log_2{k \choose k_\text{top}}\right\rceil$ bits under SLQ with $ \ell  =  \left\lceil\frac{k_\text{top}}{4(\beta_s-\delta)}\right\rceil$ will satisfy the source distortion requirement. 


\section*{Proof of Lemma \ref{lma5}}
We can bound the end-to-end expected distortion as
\begin{align}
 \mathbb{E}[D_\text{TV}(\mathbf{p},\hat{\mathbf{p}}(\kappa(\mathbf{y}))] 
 & \hspace{-0.1cm} \stackrel{(a)}{=} \hspace{-0.1cm} P(\psi(\mathbf{p}) {=} \kappa(\mathbf{y}))D_1
\nonumber  \hspace{-0.1cm} +  \hspace{-0.1cm} P(\psi(\mathbf{p}) \neq  \kappa(\mathbf{y})) D_2 \\
&\stackrel{(b)}{\leq} (1-\epsilon^{*}(n)) D_1 + \epsilon^{*}(n)D_2
\nonumber \\
&\stackrel{(c)}{\leq} (1-\epsilon^{*}(n)) \beta_s + \epsilon^{*}(n),
\end{align}
where (a) follows from the total probability theorem, and $D_\text{1} = \mathbb{E}[D_\text{TV}(\mathbf{p},\psi(\mathbf{p}))| \psi(\mathbf{p})=\kappa(\mathbf{y})]$ ($D_\text{2} = \mathbb{E}[D_\text{TV}(\mathbf{p},\hat{\mathbf{p}})|\psi(\mathbf{p}) \neq \kappa(\mathbf{y})]$, respectively) is the expected distortion when quantized probability vector is exactly constructed (not exactly reconstructed, respectively) at the receiver;   
(b) follows from considering a bound on the decoding error probability such that, $\mathbb{P}(\psi(\mathbf{p}) \neq  \kappa(\mathbf{y})) \leq \epsilon^{*}(n)$. 
(c) follows using the source distortion constraint $D_1 \leq \beta_s$, and from the fact that the total distortion is quantified using TV-divergence which allows us to bound $D_2 \leq 1$. 


%

\newpage

 




\vfill

\end{document}